\documentclass{article}

\usepackage[preprint, nonatbib]{neurips_2024}

\usepackage[utf8]{inputenc} 
\usepackage[T1]{fontenc}    
\usepackage{hyperref}       
\usepackage{url}            
\usepackage{booktabs}       
\usepackage{amsfonts}       
\usepackage{nicefrac}       
\usepackage{microtype}      
\usepackage{xcolor}         
\usepackage{graphicx} 
\usepackage{titlesec}
\usepackage{enumitem}
\usepackage{multirow}
\usepackage{tabularx, colortbl}
\usepackage{makecell}
\usepackage{caption} 
\setitemize{leftmargin=4mm}

\usepackage{amsmath,amsfonts,bm}

\def\eqref#1{equation~\ref{#1}}

\def\1{\bm{1}}

\def\ve{{\bm{e}}}

\def\vh{{\bm{h}}}

\def\vs{{\bm{s}}}

\def\vx{{\bm{x}}}

\def\mA{{\bm{A}}}

\def\mE{{\bm{E}}}

\def\mW{{\bm{W}}}
\def\mX{{\bm{X}}}

\DeclareMathAlphabet{\mathsfit}{\encodingdefault}{\sfdefault}{m}{sl}
\SetMathAlphabet{\mathsfit}{bold}{\encodingdefault}{\sfdefault}{bx}{n}

\newcommand{\R}{\mathbb{R}}

\newcommand{\softmax}{\mathrm{softmax}}

\definecolor{Gray}{gray}{0.9}

\hypersetup{
    colorlinks,
    linkcolor={blue!0!black},
    citecolor={blue!80!black},
    urlcolor={blue!80!black}
}

\title{BrainMAE: A Region-aware Self-supervised Learning Framework for Brain Signals}

\author{%
  Yifan Yang$^1$,  ~Yutong Mao$^1$, ~Xufu Liu$^1$, ~Xiao Liu$^{1,2}$\\
  $^1$Department of Biomedical Engineering, The Pennsylvania State University\\
  $^2$Institute for Computational and Data Sciences, The Pennsylvania State University\\
  \texttt{\{yzy161, yzm5278, xbl5292, xxl213\}@psu.edu} \\
}

\begin{document}

\maketitle

\begin{abstract}
The human brain is a complex, dynamic network, which is commonly studied using functional magnetic resonance imaging (fMRI) and modeled as network of Regions of interest (ROIs) for understanding various brain functions. Recent studies utilize deep learning approaches to learn the brain network representation based on functional connectivity (FC) profile, broadly falling into two main categories. The Fixed-FC approaches, utilizing the FC profile which represents the linear temporal relation within the brain network, are limited by failing to capture informative brain temporal dynamics. On the other hand, the Dynamic-FC approaches, modeling the evolving FC profile over time, often exhibit less satisfactory performance due to challenges in handling the inherent noisy nature of fMRI data. 

To address these challenges, we propose Brain Masked Auto-Encoder (BrainMAE) for learning representations directly from fMRI time-series data. Our approach incorporates two essential components—a region-aware graph attention mechanism designed to capture the relationships between different brain ROIs, and a novel self-supervised masked autoencoding framework for effective model pre-training. These components enable the model to capture rich temporal dynamics of brain activity while maintaining resilience to inherent noise in fMRI data. Our experiments demonstrate that BrainMAE consistently outperforms established baseline methods by significant margins in four distinct downstream tasks. Finally, leveraging the model's inherent interpretability, our analysis of model-generated representations reveals findings that resonate with ongoing research in the field of neuroscience. The code is available at \url{https://anonymous.4open.science/r/fMRI-State-F014}.
\end{abstract}

\setlength{\abovedisplayskip}{4pt} \setlength{\abovedisplayshortskip}{4pt}
\setlength{\belowdisplayskip}{4pt} \setlength{\belowdisplayshortskip}{4pt}

\titlespacing*{\section} {0pt}{0.5ex}{0.5ex}
\titlespacing*{\subsection} {0pt}{0.1ex}{0.1ex}
\titlespacing*{\subsubsection} {0pt}{0.1ex}{0.1ex}

\section{Introduction}
Functional magnetic resonance imaging (fMRI) is a non-invasive neuroimaging technique used to measure brain activity. Due to its high spatial and temporal resolution, fMRI has become a cornerstone of neuroscience research, enabling the study of brain functions \cite{bullmore2009complex, bassett2017network}. A general practice is to extract some brain region-of-interests (ROIs) and conceptualize the brain as a network composed of these ROIs. The connectivity between these ROIs is defined based on their linear temporal relationships, i.e. the correlation between the ROI signals. This profile of functional connectivity (FC) serves as a valuable biomarker, offering insights into the study of brain diseases \cite{greicius2008resting, wang2007altered}, aging \cite{ferreira2013resting, dennis2014functional}, and behaviors \cite{smith2015positive}, and has emerged as a key tool for understanding brain function.

Recent advances have sought to leverage the rich information contained in fMRI data by applying deep learning techniques, capitalizing on their capacity for high-level representation learning. The prevalent approach in this domain involves employing Graph Neural Networks (GNNs) to extract intricate brain network representations, which can then be applied to tasks such as decoding human traits or diagnosing diseases \cite{kan2022brain, kawahara2017brainnetcnn, kim2021learning}. These models can be broadly classified into two categories based on their treatment of temporal dynamics within the data. The first category, referred to as Fixed-FC models, relies on FC matrices computed from the entire time series of fMRI data. In contrast, the second category, known as Dynamic-FC models, takes into account the temporal evolution of brain networks. These models compute FC within temporal windows using a sliding-window approach or directly learn FC patterns from the time-series data \cite{kan2022fbnetgen}. However, both types of models exhibit certain limitations, primarily due to the unique characteristics of fMRI data.

For Fixed-FC models, depending solely on the FC profile can limit their representational capacity, as it overlooks the valuable temporal dynamics inherent in brain activities. These dynamics are often considered essential for capturing the brain's evolving states, and failing to account for them results in suboptimal brain network representations \cite{hutchison2013dynamic, preti2017dynamic, liu2018subcortical}. However, current Dynamic-FC models that consider dynamic properties, often underperform Fixed-FC models \cite{kim2021learning}. This discrepancy can be attributed to the intrinsic noisy nature of fMRI signals, as modeling temporal dynamics may amplify noise to some extent, whereas Fixed-FC approaches tend to mitigate noise by summarizing FC matrices using the entire time series \cite{hutchison2013dynamic, vidaurre2019stable, alonso2023toward}. Furthermore, FC has been shown to be sensitive to denoising preprocessing pipelines in neuroscience studies, potentially limiting the generalizability of model representations to differently preprocessed fMRI data \cite{parkes2018evaluation, li2019global, van2012influence}.

In response to these challenges, we propose Brain Masked Auto-Encoder (BrainMAE), a novel approach for learning representations from fMRI data. Our approach captures the rich temporal dynamics present in fMRI data and mitigates the impact of inherent noise through two essential components. First, drawing inspiration from the practice of word embeddings in natural language processing (NLP) \cite{devlin2018bert}, we maintain a shared embedding vector for each brain ROI. These ROI embeddings are learned globally using fMRI data from all individuals in the dataset, enabling us to obtain rich and robust representations of each ROI. Based on these ROI embeddings, we introduce a region-aware attention mechanism, adhering to the constrained nature of brain network connectivity, thus providing a valuable constraint in feature learning. Second, we leverage the information contained within the fMRI data by introducing a novel pretraining framework inspired by the concept of masked autoencoding in NLP and computer vision research \cite{brown2020language, he2022masked}. This masked autoencoding approach empowers the model to acquire genuine and transferable representations of fMRI time-series data. By integrating these two components, BrainMAE consistently outperforms existing models by significant margins across several distinct downstream tasks. Furthermore, owing to its transformer-based design and inclusion of temporal components, BrainMAE provides interpretable results, shedding light on the insights it learns from the data. Lastly, we evaluate the model-generated representations, revealing intriguing findings that align with ongoing research in the field of neuroscience.

\section{Approach}
Our approach incorporates two essential components: a novel region-aware graph attention mechanism and a masked autoencoding pretraining framework tailored for fMRI representation learning.
\subsection{Region-aware Graph Attention}
\label{three-roi-properties}
We motivate our region-aware attention module based on the inherent characteristics of brain ROIs.
\begin{itemize}[nosep]
    \item \textbf{Functional Specificity.} The brain is organized as a distributed system, with each distinct brain region serving a specific and well-defined role in the overall functioning of the brain \cite{power2011functional}.

    \item \textbf{Functional Connectivity.} Different brain regions often are interconnected and collaborate to facilitate complex cognitive functions \cite{bassett2017network}.

    \item \textbf{Inter-Individual Consistency.} Brain regions are known to exhibit a relatively consistent functional profile across different individuals. For instance, the primary visual cortex consistently processes visual information in nearly all individuals \cite{kliemann2019intrinsic}.
\end{itemize}

\begin{figure*}[t]
\begin{center}
\includegraphics[width=0.96\textwidth]{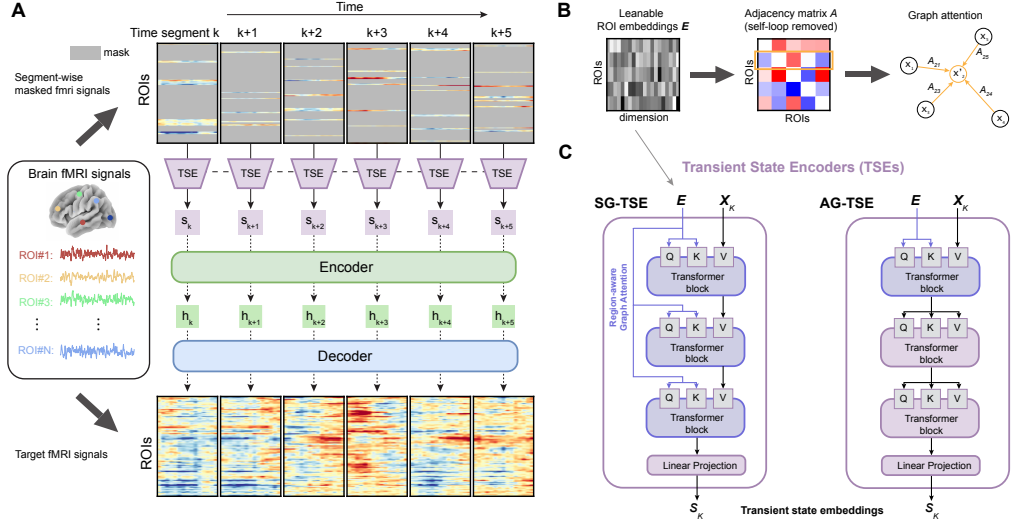}
\end{center}
\vspace{-3mm}
\caption{Overview of the Proposed BrainMAE Method. (A). Overall pre-training procedures for BrainMAE. (B). Region-aware Graph Attention. (C). Architecture of proposed TSE modules.}
\label{fig-1}
\vspace{-4mm}
\end{figure*}

\textbf{ROI Embeddings.} There is a notable similarity in the representation properties between brain ROIs and words in language. Every ROI and word carries specific functional meanings, and when combined into brain networks or sentences, they form more complicated concepts. Furthermore, the functional meaning of ROIs and words typically exhibit stability among different human individuals or across sentences. Therefore, motivated by language modeling research, we assign a learnable $d$-dimensional vector, referred to as ROI embedding, to each brain ROI. Then $N$ ROIs that cover the entire brain cortical regions form an embedding matrix denoted as $\mE\in \R^{N\times d}$.

\textbf{Region-aware Graph Attention Module.}
Brain functions as a network, with brain ROIs essentially interconnected to form a functional graph. Within this graph, each ROI is considered a node, with node feature represented by $\vx\in\R^{d}$ (Note: the ROI time-series signal of length $\tau$ is first projected into $d$-dimensional space). The set of nodes in the graph is denoted as $\mathcal{V}$.

Brain ROI activities are intrinsically governed by both structural and functional networks. ROIs that are \emph{functionally similar} tend to collaborate and exhibit synchronized activities \cite{bassett2017network}. Drawing from these biological insights, and considering that functional relevance between ROIs can be quantified by the similarity in embeddings, we define the embedding-based graph adjacency matrix $\mA\in\R^{N\times N}$. As illustrated in Figure~\ref{fig-1}B, each entry contains the edge weight between nodes $i,j\in\mathcal{V}$:
\begin{align}
    \mA_{ij} = s(\mW_q\ve_i, \mW_k\ve_j).
\end{align}
In this equation, $s: \R^d\times\R^d \longrightarrow \R$ is the similarity measurement of two vectors, e.g., scaled dot-product and $\ve_i$, $\ve_j$ are the embedding vectors for node $i$ and $j$ respectively. The weight matrices $\mW_q, \mW_k\in\R^{d\times d}$ are learnable and introduce asymmetry into $\mA$, representing the asymmetric information flow between two ROIs. Then, adopting the idea of graph attention mechanism, we derive the attention weight between node $i$ and node $j\in\mathcal{V}\backslash \{i\}$ as:
\begin{align}
    \alpha_{ij} = \softmax_j(\mA_{i}) = \frac{\exp(\mA_{ij})}{\sum_{k\in\mathcal{V}\backslash \{i\}} \exp(\mA_{ik})}.
\end{align}
Grounded in the synchronized nature among functionally related ROIs, self-loops are removed from the attention. This design prevents the attention from favoring its own node, enhancing reliability of feature learning by aggregating information from functionally connected nodes, thus reducing its sensitivity to input noise of each node. Hence the feature extracted by the module for node $i$ is
\begin{align}
    \vx^\prime_i = \sum\nolimits_{j\in\mathcal{V}\backslash \{i\}} \alpha_{ij}\vx_j
\end{align}

For implementation, we integrate the region-aware graph attention into the standard transformer block \cite{vaswani2017attention}, employing ROI embeddings as both key and query, and using input node feature as value. 

\subsection{Brain Masked AutoEncoder}
In order to effectively capture the temporal dynamics and extract the genuine representation from fMRI signals, we utilize a transformer-based encoder-decoder architecture and design a novel nontrivial self-supervised task for pretraining, as shown in Figure~\ref{fig-1}A.

\textbf{Temporal Segmentation.}
Similar to previous studies on vision transformers, where 2D images are divided into non-overlapping patches \cite{he2022masked}, we temporally segment the fMRI signals. Each fMRI segment has the shape of $N\times\tau$, where $\tau$ represents the length of segment. Such segmentation allows transformer-like models to be seamlessly applied, as each fMRI segment can be viewed as a token, and thus the original fMRI data can be represented as a sequence of tokens. Throughout our study, $\tau$ is set to 15 seconds aligning with the typical duration of transient events in fMRI data \cite{shine2016dynamics, bolt2022parsimonious}.

\textbf{Transient State Encoders (TSEs).}
We embed each fMRI segment denoted as $\mathbf{X}_k\in\R^{N\times \tau}$ using Transient State Encoders (TSEs) with detailed architecture illustrated in Figure~\ref{fig-1}C. We introduce two types of TSEs, namely Static-Graph TSE (\textbf{SG-TSE}) and Adaptive-Graph TSE (\textbf{AG-TSE}), indicating how node features are learned.

Both TSEs consist of three transformer blocks but differ in the attention mechanisms applied within these layers. For SG-TSE, all three transformer blocks exclusively employ region-aware graph attention mechanism, assuming ``static" connectivity among brain regions. On the other hand, AG-TSE incorporates two self-attention blocks stacked on top of a region-aware attention block, allowing the attention to be ``adaptive" to the node input signal, enabling the model to capture the transient reorganization of brain connectivity.
The output from the transformer blocks forms a matrix $\mX^o\in\R^{N\times d}$, where each row represents the feature learned for each ROI. We employ a linear projection $g: \R^{N\times d} \longrightarrow \R^d$ to aggregate all of the ROI features into a single vector $s_k\in\R^{d}$. This vector, as the output of TSE, represents the transient state of fMRI segment $k$.

\textbf{Segment-wise ROI Masking.} 
Different from the masked autoencoding commonly used in image or language modeling studies \cite{he2022masked, devlin2018bert}, where tokens or image patches are typically masked out, we employ a segment-wise ROI masking approach. Specifically, for each fMRI segment, we randomly choose a subset of the ROIs, such as 70\% of the ROIs, and then mask out selected ROI signals within that segment. The masked ROI segments are replaced with a masked token, a shared and learnable $d$-dimensional vector to indicate the presence of missing ROI signals. This masking scheme introduces a nontrivial reconstruction task, guiding the model to learn functional relationships between ROIs.

\textbf{Autoencoder.}
We employ a transformer-based autoencoder to capture both the temporal relationships between fMRI segments and extract the overall fMRI representation. The encoder maps the input sequence of transient state embeddings generated by the TSE ($\vs_1, \vs_2 …, \vs_n$) to a sequence of hidden representations ($\vh_1, \vh_2, …, \vh_n$). Based on these hidden representations, the decoder reconstructs the fMRI segments ($\hat{\mX}_1, \hat{\mX}_2, ..., \hat{\mX}_n$). Both the encoder and decoder consist of two standard transformer blocks and position embeddings are added for all tokens in both the encoder and decoder. The decoder is only used in pre-training phase and omitted from downstream task fine-tuning. 

\textbf{Reconstruction Loss.}
We compute Mean Squared Error (MSE) loss to evaluate the reconstruction error for masked ROI segments and unmasked ROI segments separately:
\begin{align}
    \mathcal{L}_\text{mask} &= \sum^n_{k=1}\frac{1}{n\tau|\Omega_k|}\sum_{i\in\Omega_k}\|\hat{\mX}_{k,i} - \mX_{k,i}\|^2 \\
    \mathcal{L}_\text{unmask} &= \sum^n_{k=1}\frac{1}{n\tau|\mathcal{V}\backslash\Omega_k|}\sum_{i\in\mathcal{V}\backslash\Omega_k}\|\hat{\mX}_{k,i} - \mX_{k,i}\|^2
\end{align}
where $n$ is total number of fMRI segments, $\Omega_k$ is the set of masked ROI in the $k$-th segments, $\mX_{k,i}$ is the $k$-th fMRI segment of the $i$-th ROI and $\hat{\mX}_{k,i}$ is the reconstructed one. The total reconstruction loss for pretraining the model is the weighted sum of the two:
\begin{align}
    \mathcal{L} = \lambda \mathcal{L}_\text{mask} + (1 - \lambda) \mathcal{L}_\text{unmask}
\end{align}
where $\lambda$ is a hyperparameter, and in our study, we set $\lambda$ to a fixed value of 0.75 to penalize more on the reconstruction loss of masked ROI segments.

\textbf{BrainMAEs.}
Based on the choice of TSE, we introduce two variants of BrainMAE: SG-BrainMAE and AG-BrainMAE, incorporating SG-TSE and AG-TSE for transient state encoding, respectively.

\begin{figure*}[t]
\centering
\includegraphics[width=1\textwidth]{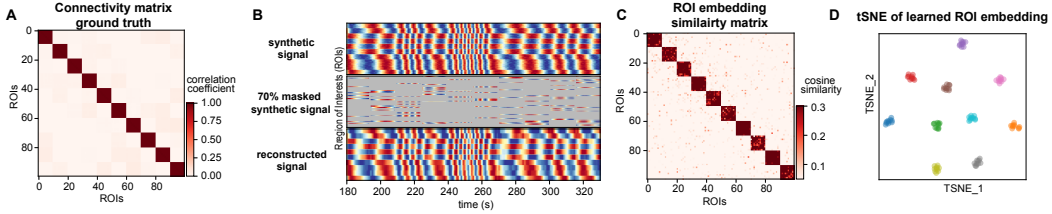}
\caption{Validation on synthetic dataset. (A) Ground truth connectivity matrix. (B) Example synthetic signals and their reconstruction using pre-trained BrainMAE. (C) Similarity matrix between learned ROI embeddings. (D) t-SNE visualization of the ROI embeddings revealing the real network.}
\label{fig-2}
\vspace{-4mm}
\end{figure*}

\section{Experiments}
\subsection{Model Validation with Synthetic Data}
We posit that our model design and tailored self-supervised pre-training scheme facilitate the learning of region-specific information within the ROI embeddings. To test this claim, we conduct a simulation study where SG-BrainMAE is pre-trained using a synthetic dataset. We pre-define a connectivity matrix among ROIs (Figure~\ref{fig-2}A) and generate synthetic signals according to the criterion that signals between two ROIs with high connectivity tend to exhibit similar fluctuation (Figure~\ref{fig-2}B; See Appendix~\ref{appdx-simulation-analysis} for detailed simulation setup). As shown in Figure~\ref{fig-2}C and \ref{fig-2}D, the similarity matrix of ROI embedding trained with this synthetic dataset, accurately captures the ground truth connectivity. Furthermore, the t-SNE plot of ROI embeddings demonstrates that ROIs with strong connectivity are correctly closely clustered, thereby providing validation for our claim and design principles. 
\subsection{fMRI Datasets}
We mainly use the following fMRI datasets to evaluate our approach.
\textbf{HCP-3T} dataset \cite{van2013wu} is a large-scale publicly available dataset that includes 3T fMRI data from 897 healthy adults aged between 22 and 35. We use both the resting-state and task sessions as well as the behavior measurements in our study. 
\textbf{HCP-7T} dataset is a subset of the HCP S1200 release, consisting of 7T fMRI data from 184 subjects within the age range of 22 to 35. Our analysis focused on the resting-state sessions of this dataset.
\textbf{HCP-Aging} dataset \cite{harms2018extending}, designed for aging studies, contains 725 subjects aged 36 to 100+. Age, gender information as well as the resting-state fMRI data are used in our study.
\textbf{NSD} dataset \cite{allen2022massive} is a massive 7T fMRI dataset, featuring 8 human participants, each with 30-40 scan sessions conducted over the course of a year. For our study, we incorporated both task fMRI data and task performance metrics, including task scores and response times (RT).
Detailed information regarding each of the datasets can be found in Appendix \ref{appdx-dataset}.

\begin{figure*}[t]
\begin{center}
\includegraphics[width=0.95\textwidth]{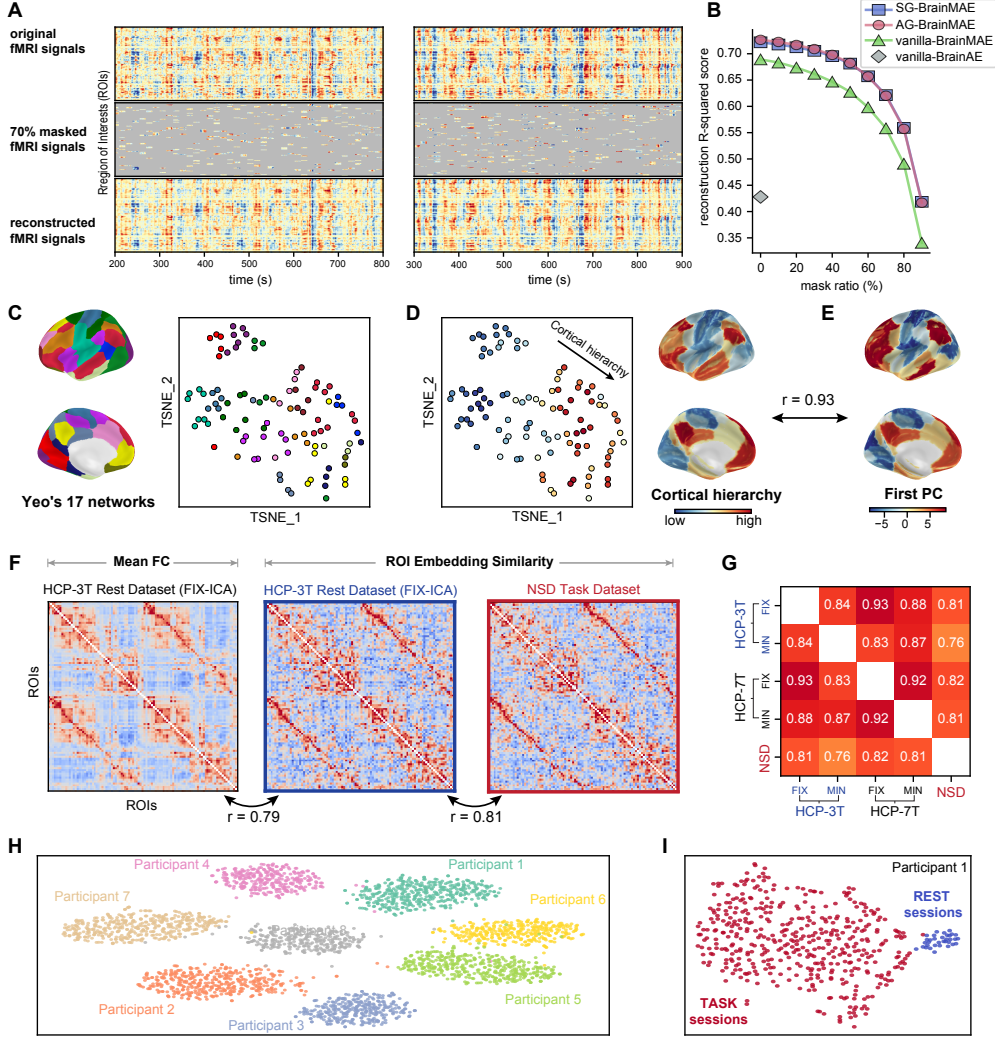}
\end{center}
\vspace{-2mm}
\caption{Evaluation of pre-trained BrainMAE. (A) Reconstruction examples on unseen fMRI datasets. (B) Reconstruction performance across various mask ratios. (C-D) t-SNE plot of the ROI embeddings pre-trained with HCP-3T Rest Dataset, revealing Yeo-17 network (C) and brain cortical hierarchy (D). (E) First principal component of ROI embeddings aligns with the cortical hierarchy. (F-G) ROI embedding similarity matrix shows high degrees of similarity to the FC matrix (F) and is consistent across models pretrained with different datasets (G). (H-I) t-SNE plot of fMRI representation ($h_{[CLS]}$) on unseen NSD dataset to reveal subject identity (H) and differentiate between task and rest states (I).}
\label{fig-3}
\vspace{-4mm}
\end{figure*}

\subsection{Pre-training Evaluation}
\subsubsection{Implementation Details}
During the pretraining phase, each time we randomly select 300 seconds of fixed-length fMRI signals from the original sample and divide this signal into 20 segments, with each segment containing 15 seconds of fMRI data. We use a variable mask ratio for each mini-batch during training. The mask ratio for each mini-batch is drawn from a range of (0, 0.8), where 0 indicates no masking is applied. For each pretrained dataset, we train the model using all available samples for 1000 epochs. Additional training settings are available in Appendix~\ref{appdx-pretrain-setting}.

\subsubsection{Masked Signal Reconstruction}

We assess the reconstruction performance of the HCP-3T pretrained model on unseen HCP-7T dataset. Example reconstruction results are shown in Figure~\ref{fig-3}A and Appendix \ref{appdx-more_recon}. Figure~\ref{fig-3}B presents comparative analysis of the reconstruction performance, quantified by $R^2$ values, between the proposed models and variants employing only self-attention within the TSE module. These variants are denoted as vanilla-BrainMAE (pretrained using the proposed masked autoencoding, see \ref{appdx-fdgmae}) and vanilla-BrainAE (pretrained with standard autoencoding, see \ref{appdx-fdgae}). Both SG-BrainMAE and AG-BrainMAE achieve better reconstruction performance across all mask ratios, suggesting the incorporation of region-aware graph attention is advantageous for learning generalized representations.

\subsubsection{ROI Embeddings}
It is crucial to validate whether the ROI embeddings pretrained with the proposed approach truly satisfy the aforementioned three ROI characteristics in section~\ref{three-roi-properties}. 

\textbf{Functional Specificity.} We visualize t-SNE transformed ROI embeddings in Figure~\ref{fig-3}C \cite{van2008visualizing}. In the projected 2D space, the ROIs exhibit discernible clustering that aligns with the Yeo-17 networks' definitions in neuroscience studies \cite{yeo2011organization}. This alignment suggests that ROIs with similar functional roles (e.g., visual processing) display similar embeddings. In other words, specific brain functions can be inferred from these embeddings.

\begin{table*}
\setlength{\tabcolsep}{4 pt}
\renewcommand{\arraystretch}{1}
\begin{center}
\caption{Results for behavior prediction.}
\label{tbl-1}
\centering
\resizebox{1\linewidth}{!}{
\begin{tabular}{llllllllllll}
\toprule
\multirow{2.5}{*}{Model}&\multicolumn{2}{c}{Gender}&&\multicolumn{8}{c}{Behaviors (measured in MAE)}\\
\cmidrule(lr){2-3} \cmidrule(lr){5-12} 
&{Accuracy (\%)}&{AUROC (\%)}&{}&{PicSeq}&{PMT$\_$CR}&{PMT$\_$SI}&{PicVocab}&{IWRD}&{ListSort}&{LifeSatisf}&{PSQI} \
\\ \hline 
\textit{\textbf {Fixed-FC}}\\
BrainNetTF-OCR & 94.11\small{±0.98} & 94.39\small{±0.90} &{}& 7.11\small{±0.22} & 2.28\small{±0.11} & 1.72\small{±0.12} & 4.70\small{±0.10} & 1.56\small{±0.03} & 5.96\small{±0.07}& 4.96\small{±0.22}& 1.49\small{±0.05}\\
BrainNetTF-Vanilla & 90.00\small{±1.05} & 89.93\small{±0.94} &{}& 8.19\small{±0.46} & 2.73\small{±0.07} & 2.13\small{±0.07} & 5.93\small{±0.14} & 1.81\small{±0.08} & 6.91\small{±0.21} & 5.71\small{±0.08} & 1.68\small{±0.05} \\
BrainNetCNN & 90.68\small{±1.80} & 90.89\small{±1.55} &{}& 10.21\small{±0.22} & 3.25\small{±0.13} & 2.64\small{±0.14} & 6.65\small{±0.27} & 2.26\small{±0.05} & 8.51\small{±0.20} & 7.12\small{±0.24} & 1.68\small{±0.05} \\
 \hline 
\textit{\textbf {Dynamic-FC}}\\
STAGIN-SERO & 88.73\small{±1.36} & 88.69\small{±1.41} &{}& 10.22\small{±0.15} & 3.49\small{±0.05} & 2.70\small{±0.06}& 6.78\small{±0.20} & 2.26\small{±0.05} & 8.51\small{±0.20} & 7.12\small{±0.24} & 2.12\small{±0.04}\\
STAGIN-GARO & 88.34\small{±0.94} & 88.33\small{±0.91} &{}& 10.26\small{±0.18} & 3.44\small{±0.10} & 2.69\small{±0.09} & 6.92\small{±0.30} & 2.25\small{±0.04} & 8.52\small{±0.26} & 7.09\small{±0.35} & 2.08\small{±0.04} \\
FBNETGNN & 88.05\small{±0.15} & 87.93\small{±0.97} &{}& 8.62\small{±0.21} & 2.93\small{±0.11} & 2.34\small{±0.11} & 5.83\small{±0.15} & 1.96\small{±0.04} & 7.31\small{±0.10} & 6.09\small{±0.10} & 1.81\small{±0.03} \\
\hline 
\textit{\textbf {Ours}}\\
vanilla-BrainAE & 94.11\small{±1.02} & 94.07\small{±1.09} &{}& 7.63\small{±0.27} & 2.50\small{±0.12} & 1.92\small{±0.11} & 5.01\small{±0.17} & 1.67\small{±0.04} & 6.45\small{±0.28} & 5.40\small{±0.13} & 1.59\small{±0.05} \\
vanilla-BrainMAE & 95.80\small{±1.23} & 96.13\small{±1.09} &{}& 5.11\small{±0.15} & 1.69\small{±0.06} & 1.30\small{±0.07} & 3.40\small{±0.11} & 1.12\small{±0.05} & \bf 4.33\small{±0.16} & 3.60\small{±0.15} & \bf 1.05\small{±0.06}\\
\rowcolor{Gray}
SG-BrainMAE & \bf 97.49\small{±0.15} & \bf 97.46\small{±0.18} &{}& \bf 5.06\small{±0.21} & \bf 1.63\small{±0.08} & \bf 1.24\small{±0.04} & 3.40\small{±0.14} & \bf 1.11\small{±0.04} & 4.35\small{±0.12} &  3.64\small{±0.27} & \bf 1.05\small{±0.06}\\
\rowcolor{Gray}
AG-BrainMAE &  97.13\small{±0.56} &  97.17\small{±0.61} &{}&  5.09\small{±0.05} &  1.67\small{±0.10} &  1.28\small{±0.06} &  \bf 3.34\small{±0.11} &  1.13\small{±0.03} &  4.37\small{±0.06} &  \bf 3.58\small{±0.17} &  1.07\small{±0.05} \\
\bottomrule
\end{tabular}
}
\end{center}
\vspace{-8mm}
\end{table*}

\begin{table}
  \parbox{.51\linewidth}{
  \centering
  \caption{Results for age prediction.}
  \label{tbl-2}
  \resizebox{1\linewidth}{!}{
  \begin{tabular}{lcccc}
    \toprule
    \multirow{2.5}{*}{Model}&\multicolumn{2}{c}{ Gender}&{}&\multirow{2.5}{*}{ Age (MAE)}\\
\cmidrule(lr){2-3} 
&{Accuracy (\%)}&{AUROC (\%)}
\\ \hline 
\textit{\textbf {Fixed-FC}}\\
BrainNetTF-OCR & 90.21{\small±3.81} & 90.73{\small±2.85} &{}& 6.15{\small±0.71} \\
BrainNetTF-Vanilla & 88.96{\small±2.16} & 88.76{\small±2.21} &{}& 6.78{\small±0.56}  \\
BrainNetCNN & 88.83{\small±1.52} & 88.74{\small±1.58} &{}& 8.71{\small±0.62}  \\
 \hline 
\textit{\textbf {Dynamic-FC}}\\
STAGIN-SERO & 82.37\small{±1.66} & 82.57\small{±1.36} &{}& 8.96\small{±0.47} \\
STAGIN-GARO & 80.67\small{±0.81} & 80.58\small{±1.03} &{}& 8.65\small{±0.28}  \\
FBNETGNN & 89.50\small{±3.58} & 89.34\small{±3.49} &{}& 6.68\small{±1.00} \\
\hline 
\textit{\textbf {Ours}}\\
vanilla-BrainAE & 80.92\small{±2.40} & 81.03\small{±2.52} &{}& 8.33\small{±0.49}  \\
vanilla-BrainMAE & 88.54\small{±2.50} & 88.53\small{±2.37} &{}& 7.26\small{±0.95}  \\
\rowcolor{Gray}
SG-BrainMAE & \bf 92.67\small{±1.07} & \bf 92.51\small{±1.07} &{}& \bf 5.78\small{±0.44} \\
\rowcolor{Gray}
AG-BrainMAE &  91.12\small{±1.99} &  91.15\small{±2.03} &{}&  6.49\small{±1.00} \\
    \bottomrule
\end{tabular}
}
}
\hfill
\parbox{0.48\linewidth}{
  \centering
  \caption{Results for task performance prediction.}
  \label{tbl-3}
  \resizebox{1\linewidth}{!}{
\begin{tabular}{lcccc}
\toprule
\multirow{2.5}{*}{Model}&{}&\multirow{2.5}{*}{Task score (MAE)}&{}&\multirow{2.5}{*}{RT (MAE in ms)}\\
\addlinespace
\\ \hline 
\textit{\textbf {Fixed-FC}}\\
BrainNetTF-OCR&{}&0.070\small{±0.003}&{}&92.344\small{±2.343}  \\
BrainNetTF-Vanilla &{}& 0.075\small{±0.004}&{}&96.252\small{±2.133}   \\
BrainNetCNN &{}& 0.078\small{±0.004}&{}&102.911\small{±2.225} \\
 \hline 
\textit{\textbf {Dynamic-FC}}\\
STAGIN-SERO&{}& 0.089\small{±0.003} &{}&116.635\small{±2.197} \\
STAGIN-GARO &{}& 0.091\small{±0.002} &{}& 116.130\small{±2.099}   \\
FBNETGNN &{}& 0.074\small{±0.005} &{}& 95.349\small{±2.320}  \\
\hline 
\textit{\textbf {Ours}}\\
vanilla-BrainAE  &{}& 0.091\small{±0.004} &{}& 118.965\small{±3.047} \\
vanilla-BrainMAE &{}& 0.083\small{±0.004} &{}& 108.215\small{±3.458}   \\
\rowcolor{Gray}
SG-BrainMAE &{}& \bf 0.069\small{±0.004} &{}& \bf 90.678\small{±1.767}  \\
\rowcolor{Gray}
AG-BrainMAE &{}&  0.070\small{±0.003} &{}&  92.154\small{±2.265} \\
\bottomrule
    \end{tabular}
}
}
\vspace{-4mm}
\end{table}

\textbf{Functional Connectivity.} Interestingly, the arrangement of the ROI in the projected 2D space also reflects the cortical hierarchy, as indicated by principal gradient (PG) values (Figure~\ref{fig-3}D-E) \cite{margulies2016situating, gu2021brain, raut2021global}. Low PG values correspond to cortical low-order regions, such as visual and somatosensory regions, while high PG values correspond to cortical high-order regions, including the default mode network and limbic system. Therefore, the interconnectivity between different brain networks, captured by the PG, thus can also be informed by the ROI embeddings.

\textbf{Inter-Individual Consistency.} We separately pretrain the SG-BrainMAE on the HCP-3T, HCP-7T, and NSD task datasets. Both HCP-3T and HCP-7T datasets have two different preprocessing pipelines, namely minimal preprocessing and FIX-ICA. Consequently, we pretrain models for each combination. In total, we obtain five independently pretrained models. For each pretrained model, we generate an embedding similarity matrix by computing pairwise cosine similarities between ROIs embeddings, as shown in Figure~\ref{fig-3}F-G. Importantly, these similarity matrices exhibit consistent patterns across different datasets, regardless of preprocessing pipelines or fMRI task types (resting or task), suggesting the converging ROI representations. 

Despite general consistency of ROI embeddings across datasets, the modular structure of network constructed based on embeddings similarity matrix, quantified by modularity, shows a reduction with aging (Figure~\ref{appdx-fig-9}), aligning with the established findings in neuroscience research \cite{sporns2016modular, wig2017segregated}. More ROI analysis is shown in Appendix \ref{appdx-emb-const}.

\subsubsection{Representation Analysis}
To study pretrained fMRI representation (output of CLS token $h_{[CLS]}$), we evaluate the HCP-3T pretrained model on unseen NSD dataset. Notably, as shown in Figure~\ref{fig-3}H and \ref{fig-3}I, the t-SNE plot of the fMRI representation demonstrates clear separation between fMRI sessions from different subjects. Within individuals, resting-state fMRI sessions are distinctly separated from task-based fMRI sessions. These results suggest that the pre-trained fMRI representation carries valuable information for distinguishing individuals and reflecting their respective brain arousal states.

\subsection{Transfer Learning Evaluation}
\subsubsection{Implementation Details}
We pretrain the model using HCP-3T ICA-FIX preprocessed data and fine-tune the whole network except for the ROI embeddings for each downstream task. Similar to previous studies \cite{he2022masked, devlin2018bert}, we use the CLS token from the encoder and append a task-specific linear head for prediction. Cross-entropy loss and mean squared error (MSE) loss are used for classification and regression fine-tuning tasks respectively. For fair comparisons, we use 5-fold cross-validation for model evaluation. Detailed information regarding the architecture and training settings can be found in Appendix~\ref{appdx-transfer-setting}. 

\subsubsection{Steady-state Variables Prediction}
For the prediction of steady-state variables, models typically employ entire fMRI signals spanning hundreds of seconds to make predictions regarding an individual's traits, age, and gender information. We benchmark our approach against two categories of baseline neural network models specifically designed for these tasks. \textbf{(1) Baseline models based on fixed brain network (Fixed-FC).} BrainNetTF-OCR ~\cite{kan2022brain} is transformer-based model with Orthonormal Clustering Readout (OCR), representing the state-of-the-art method for brain network analysis. BrainNetTF-Vanilla is a variant of BrainNetTF with CONCAT-based readout. BrainNetCNN~\cite{kawahara2017brainnetcnn} follows the CNN paradigm by modeling the functional connectivity matrices similarly as 2D images. \textbf{(2) Baseline models based on dynamic brain network (Dynamic-FC).} STAGIN~\cite{kim2021learning} is a transformer-based model that learns dynamic graph representation with spatial-temporal attention. Two variants, STAGIN-GARO and STAGIN-SERO that use different readout methods are included for comparison. FBNETGEN~\cite{kan2022fbnetgen} is a GNN-based model learning the brain network from the fMRI time-series signals.

\begin{table}[t]
  \parbox{.48\linewidth}{
  \centering
  \caption{Results for transient mental state decoding}
  \label{tbl-mental-state-hcptask}
  \resizebox{1\linewidth}{!}{
    \begin{tabular}{lcc}
    \toprule
    Model&Accuracy(\%)&macro F1-score
    \\ \hline 
    \textit{\textbf {Self-supervised baselines}}\\
    CSM & 94.8±0.35 & 92.0   \\
    Seq-BERT & 89.20±0.35 & 84.5 \\
    Net-BERT & 89.80±0.48 & 85.2 \\
    \hline 
    \textit{\textbf {Ours}}\\
    vanilla-BrainAE & 94.24±1.06 & 93.31±1.07 \\
    vanilla-BrainMAE & 95.56±1.37 &  94.93±1.52   \\
    \rowcolor{Gray}
    SG-BrainMAE &  95.71±1.30 &  95.24±1.15  \\
    \rowcolor{Gray}
    AG-BrainMAE & \bf 95.98±1.62 & \bf 95.59±1.51 \\
    \bottomrule
    \end{tabular}
    }
}
\hfill
\parbox{.48\linewidth}{
    \centering
    \resizebox{1\linewidth}{!}{
    \includegraphics[width=0.7\textwidth]{Figures/fig4\_crop.pdf}
    }
    \vspace{-4mm}
    \captionof{figure}{Model ``pays more attention" on task blocks and the attention score correlates with brain arousal state change (measured by 1/RT).}
    \label{fig-4}
}
\vspace{-10mm}
\end{table}

We compare BrainMAEs against baseline methods across three distinct downstream tasks. \textbf{(1) Behaviors prediction.} In this task, the models are required to simultaneously perform gender classification as well as predict 8 cognitive behaviors (See Table~\ref{appdx-tbl-beh} for details) using HCP-3T dataset. \textbf{(2) Age prediction.} For this task, the models are required to simultaneously perform gender classification and age prediction on HCP-Aging dataset. \textbf{(3) Task performance prediction.} For this task, the models are required to predict the averaged memory task score as well as the averaged response time (RT) for each fMRI run using NSD dataset. 

The results for the three downstream tasks are shown in Table~\ref{tbl-1},~\ref{tbl-2}, and~\ref{tbl-3} and reported as mean $\pm$ std. from 5-fold cross-validation, with regression variables measured in mean absolute error (MAE). Our proposed methods consistently demonstrate superior performance across all these tasks. 

Notably, despite the informative nature of temporal features, the three baseline models that leverage dynamic FC or learn the FC from the fMRI time series consistently underperform compared to models that utilize fixed FC. One plausible explanation could be attributed to the inherent noise present in fMRI data \cite{kim2021learning}. In contrast, even though BrainMAEs encode the complete temporal information of fMRI, they still achieve the highest level of performance. This achievement can be attributed to the region-aware graph attention module employed in our model design.

\subsubsection{Transient Mental State Decoding}
We evaluate capability of BrainMAEs to decode transient brain states lasting only tens of seconds, a task that poses challenges for FC-based models due to the emergence of spurious connectivity with such short time window \cite{savva2019assessment}. Hence, we compare our methods with current state-of-the-art transformer-based self-supervised learning approaches tailored for fMRI transient state modeling.

\textbf{Baseline self-supervised approaches} \cite{thomas2022self}. All of the baseline methods consider each time point as word and leverage recent successes in NLP pre-training techniques to model fMRI transient states. Causal Sequence Modeling (CSM) is pretrained based on the principle of causal language modeling, predicting the next signal using historical context. Sequence-BERT performs self-supervised learning by solving masked-language-modeling and next-sentence-prediction tasks. Network-BERT, a variant to Sequence-BERT, is designed to infer the entire timeseries of a masked network of ROIs.

Following the same experimental setup as in \cite{thomas2022self}, we use publicly available HCP-3T task datasets \cite{van2013wu} and identify 20 mental states across experimental tasks. As shown in Table~\ref{tbl-mental-state-hcptask}, BrainMAEs outperform other self-supervised approaches. It's important to note that we only use 100 cortical ROIs, in contrast to baseline methods utilizing 1024 Dictionary Learning Functional Modes (DiFuMo) ROIs including subcortical areas \cite{dadi2020fine}. This comparison leads to two insights: firstly, cortical region activity alone might suffice for decoding mental states; secondly, given the highly correlated nature of fMRI signals, a small set of ROIs can effectively represent brain activity, potentially enabling the development of more efficient models for future research. Furthermore, AG-BrainMAE exhibits enhanced performance relative to SG-BrainMAE, indicating that integrating an adaptive component is beneficial for capturing transient state changes, while SG-BrainMAE is more suitable for steady-state variables prediction. Table~\ref{appdx-tbl-mental-state-hcptask-detial} presents detailed results for each mental state from the multi-class decoding task, demonstrating consistently high decoding accuracy that appears to be insensitive to the duration of each distinct mental state.

\subsubsection{Ablation Study}
We conduct ablation studies on the aforementioned four downstream tasks to elucidate the advantage of incorporating the region-aware graph attention module and masked autoencoding in our approach. We compare four model variants: (1) SG-BrainMAE, (2) AG-BrainMAE, (3) vanilla-BrainMAE (BrainMAE without the region-aware attention modules, see \ref{appdx-fdgmae} for details), and (4) vanilla-BrainAE (sharing the same architecture as vanilla-BrainMAE but pretrained with standard autoencoding, see \ref{appdx-fdgae} for details).  The differences between these variants are detailed in Table~\ref{appdx-tbl-distinction-model-variants}. The results, which are presented in Table~\ref{tbl-1},~\ref{tbl-2}, ~\ref{tbl-3} and \ref{tbl-mental-state-hcptask}, indicate a degradation in performance when the region-aware attention is removed, and a further significant decrease in performance when masked pretraining is excluded. This underscores the advantage of incorporating both components in BrainMAE framework.

\subsubsection{Interpretation Analysis}
We interpret BrrainMAE fine-tuned on NSD task performance dataset. We evaluate self-attention scores used to generate the CLS representation of the final layers of the encoder transformer. These attention scores provide insights into which time segments are crucial for the model to make predictions. As illustrated in Figure~\ref{fig-4}, we average the attention scores across different fMRI runs. Notably, the attention score reveals the NSD task block structure, indicating that the model inherently places more importance on task blocks to infer overall task performance. Interestingly, the attention score correlates to behavioral arousal measurements (inverse response time \cite{makovac2019response}), suggesting the model is aware of the change of transient brain state. Indeed, the learned fMRI representation also highly correlates with brain arousal index (Appendix~\ref{appdx-brain-state}). Overall, these results underline the interpretability of BrainMAE, hinting at its potential to explore brain mechanisms for neuroscience research. 

\section{Related Work}
\textbf{Masked autoencoding.}
The incorporation of masked autoencoding for self-supervised representation learning has seen significant interest across various domains. In the realm of NLP, models like BERT \cite{devlin2018bert} and GPT \cite{brown2020language, radford2018improving, radford2019language} employ masked autoencoding to pretrain language models by predicting the missing components of input sequences. In computer vision, masked modeling has been integrated with vision transformers, yielding successful approaches such as MAE~\cite{he2022masked}, BEiT~\cite{bao2021beit}, and BEVT~\cite{wang2022bevt} for feature learning. Limited studies explore masked autoencoding in the context of fMRI. \cite{thomas2022self} adapt BERT-like framework by considering each time point as 'word', which is suitable for modeling transient states but limited in scaling to fMRI signals of hundreds seconds. Our approach differs by treating transient state as basic unit for sequence learning, allowing it to scale effectively and extract representations that reflect individual traits and behaviors.

\textbf{Brain Network Analysis.}
GNN-based models have been widely used in the field of brain network analysis \cite{li2021braingnn, said2023neurograph, yang2022data, liu2023braintgl, zhao2022dynamic, ahmedt2021graph, li2019graph, li2020pooling, cui2022interpretable}. Models like GroupINN~\cite{yan2019groupinn} introduce the concept of node grouping to enhance interpretability and reduce model size. BrainNetCNN~\cite{kawahara2017brainnetcnn} capitalizes on the topological locality of structural brain networks for graph-based feature learning. BrainNetTF~\cite{kan2022brain}, on the other hand, is a transformer-based model that employs orthonormal clustering readout to facilitate cluster-aware graph embedding learning. STAGIN~\cite{kim2021learning} focuses on learning dynamic graph representations through spatial-temporal attention mechanisms, while FBNetGen~\cite{kan2022fbnetgen} directly learns the brain network from fMRI time-series data. In contrast to these methods, which predominantly learn representation from functional connectivity (FC), our approach can effectively incorporate valuable temporal information while mitigating the impact of fMRI's intrinsic noise through specifically designed modules.

\section{Discussion and Conclusion}
Here we propose BrainMAE for effectively learning the representation from fMRI time series data. Our approach integrates two essential components: a region-aware graph attention module and a masked self-supervised pretraining framework. These components are designed to capture temporal dynamics while mitigating the inherent noise in fMRI data. The alignment of the learned ROI embeddings with existing neuroscience knowledge, along with the improvement in transfer learning tasks, confirms the effectiveness of our design.

By providing a task-agnostic representation, BrainMAE exhibits promise for applications in the field of neuroscience. Its interpretability and ability to capture transient representations make it a valuable tool for uncovering the mechanisms and dynamics of transient state changes within the brain. 

Furthermore, our approach can be extended beyond brain fMRI data. It can be applied to various domains that can be modeled as networks of functionally meaningful nodes. For instance, it could be applied to traffic network analysis, where the nodes represent either roads or spatially defined regions.


\clearpage
\begin{thebibliography}{10}

\bibitem{ahmedt2021graph}
David Ahmedt-Aristizabal, Mohammad~Ali Armin, Simon Denman, Clinton Fookes, and Lars Petersson.
\newblock Graph-based deep learning for medical diagnosis and analysis: past, present and future.
\newblock {\em Sensors}, 21(14):4758, 2021.

\bibitem{allen2022massive}
Emily~J Allen, Ghislain St-Yves, Yihan Wu, Jesse~L Breedlove, Jacob~S Prince, Logan~T Dowdle, Matthias Nau, Brad Caron, Franco Pestilli, Ian Charest, et~al.
\newblock A massive 7t fmri dataset to bridge cognitive neuroscience and artificial intelligence.
\newblock {\em Nature neuroscience}, 25(1):116--126, 2022.

\bibitem{alonso2023toward}
Sonsoles Alonso and Diego Vidaurre.
\newblock Toward stability of dynamic fc estimates in neuroimaging and electrophysiology: Solutions and limits.
\newblock {\em Network Neuroscience}, 7(4):1389--1403, 2023.

\bibitem{bao2021beit}
Hangbo Bao, Li~Dong, Songhao Piao, and Furu Wei.
\newblock Beit: Bert pre-training of image transformers.
\newblock {\em arXiv preprint arXiv:2106.08254}, 2021.

\bibitem{bassett2017network}
Danielle~S Bassett and Olaf Sporns.
\newblock Network neuroscience.
\newblock {\em Nature neuroscience}, 20(3):353--364, 2017.

\bibitem{bolt2022parsimonious}
Taylor Bolt, Jason~S Nomi, Danilo Bzdok, Jorge~A Salas, Catie Chang, BT~Thomas~Yeo, Lucina~Q Uddin, and Shella~D Keilholz.
\newblock A parsimonious description of global functional brain organization in three spatiotemporal patterns.
\newblock {\em Nature neuroscience}, 25(8):1093--1103, 2022.

\bibitem{brown2020language}
Tom Brown, Benjamin Mann, Nick Ryder, Melanie Subbiah, Jared~D Kaplan, Prafulla Dhariwal, Arvind Neelakantan, Pranav Shyam, Girish Sastry, Amanda Askell, et~al.
\newblock Language models are few-shot learners.
\newblock {\em Advances in neural information processing systems}, 33:1877--1901, 2020.

\bibitem{bullmore2009complex}
Ed~Bullmore and Olaf Sporns.
\newblock Complex brain networks: graph theoretical analysis of structural and functional systems.
\newblock {\em Nature reviews neuroscience}, 10(3):186--198, 2009.

\bibitem{chang2016tracking}
Catie Chang, David~A Leopold, Marieke~Louise Sch{\"o}lvinck, Hendrik Mandelkow, Dante Picchioni, Xiao Liu, Frank~Q Ye, Janita~N Turchi, and Jeff~H Duyn.
\newblock Tracking brain arousal fluctuations with fmri.
\newblock {\em Proceedings of the National Academy of Sciences}, 113(16):4518--4523, 2016.

\bibitem{cui2022interpretable}
Hejie Cui, Wei Dai, Yanqiao Zhu, Xiaoxiao Li, Lifang He, and Carl Yang.
\newblock Interpretable graph neural networks for connectome-based brain disorder analysis.
\newblock In {\em International Conference on Medical Image Computing and Computer-Assisted Intervention}, pages 375--385. Springer, 2022.

\bibitem{dadi2020fine}
Kamalaker Dadi, Ga{\"e}l Varoquaux, Antonia Machlouzarides-Shalit, Krzysztof~J Gorgolewski, Demian Wassermann, Bertrand Thirion, and Arthur Mensch.
\newblock Fine-grain atlases of functional modes for fmri analysis.
\newblock {\em NeuroImage}, 221:117126, 2020.

\bibitem{dennis2014functional}
Emily~L Dennis and Paul~M Thompson.
\newblock Functional brain connectivity using fmri in aging and alzheimer’s disease.
\newblock {\em Neuropsychology review}, 24:49--62, 2014.

\bibitem{devlin2018bert}
Jacob Devlin, Ming-Wei Chang, Kenton Lee, and Kristina Toutanova.
\newblock Bert: Pre-training of deep bidirectional transformers for language understanding.
\newblock {\em arXiv preprint arXiv:1810.04805}, 2018.

\bibitem{ferreira2013resting}
Luiz~Kobuti Ferreira and Geraldo~F Busatto.
\newblock Resting-state functional connectivity in normal brain aging.
\newblock {\em Neuroscience \& Biobehavioral Reviews}, 37(3):384--400, 2013.

\bibitem{greicius2008resting}
Michael Greicius.
\newblock Resting-state functional connectivity in neuropsychiatric disorders.
\newblock {\em Current opinion in neurology}, 21(4):424--430, 2008.

\bibitem{gu2020transient}
Yameng Gu, Feng Han, Lucas~E Sainburg, and Xiao Liu.
\newblock Transient arousal modulations contribute to resting-state functional connectivity changes associated with head motion parameters.
\newblock {\em Cerebral Cortex}, 30(10):5242--5256, 2020.

\bibitem{gu2021brain}
Yameng Gu, Lucas~E Sainburg, Sizhe Kuang, Feng Han, Jack~W Williams, Yikang Liu, Nanyin Zhang, Xiang Zhang, David~A Leopold, and Xiao Liu.
\newblock Brain activity fluctuations propagate as waves traversing the cortical hierarchy.
\newblock {\em Cerebral cortex}, 31(9):3986--4005, 2021.

\bibitem{harms2018extending}
Michael~P Harms, Leah~H Somerville, Beau~M Ances, Jesper Andersson, Deanna~M Barch, Matteo Bastiani, Susan~Y Bookheimer, Timothy~B Brown, Randy~L Buckner, Gregory~C Burgess, et~al.
\newblock Extending the human connectome project across ages: Imaging protocols for the lifespan development and aging projects.
\newblock {\em Neuroimage}, 183:972--984, 2018.

\bibitem{he2022masked}
Kaiming He, Xinlei Chen, Saining Xie, Yanghao Li, Piotr Doll{\'a}r, and Ross Girshick.
\newblock Masked autoencoders are scalable vision learners.
\newblock In {\em Proceedings of the IEEE/CVF conference on computer vision and pattern recognition}, pages 16000--16009, 2022.

\bibitem{hutchison2013dynamic}
R~Matthew Hutchison, Thilo Womelsdorf, Elena~A Allen, Peter~A Bandettini, Vince~D Calhoun, Maurizio Corbetta, Stefania Della~Penna, Jeff~H Duyn, Gary~H Glover, Javier Gonzalez-Castillo, et~al.
\newblock Dynamic functional connectivity: promise, issues, and interpretations.
\newblock {\em Neuroimage}, 80:360--378, 2013.

\bibitem{kan2022fbnetgen}
Xuan Kan, Hejie Cui, Joshua Lukemire, Ying Guo, and Carl Yang.
\newblock Fbnetgen: Task-aware gnn-based fmri analysis via functional brain network generation.
\newblock In {\em International Conference on Medical Imaging with Deep Learning}, pages 618--637. PMLR, 2022.

\bibitem{kan2022brain}
Xuan Kan, Wei Dai, Hejie Cui, Zilong Zhang, Ying Guo, and Carl Yang.
\newblock Brain network transformer.
\newblock {\em Advances in Neural Information Processing Systems}, 35:25586--25599, 2022.

\bibitem{kawahara2017brainnetcnn}
Jeremy Kawahara, Colin~J Brown, Steven~P Miller, Brian~G Booth, Vann Chau, Ruth~E Grunau, Jill~G Zwicker, and Ghassan Hamarneh.
\newblock Brainnetcnn: Convolutional neural networks for brain networks; towards predicting neurodevelopment.
\newblock {\em NeuroImage}, 146:1038--1049, 2017.

\bibitem{kim2021learning}
Byung-Hoon Kim, Jong~Chul Ye, and Jae-Jin Kim.
\newblock Learning dynamic graph representation of brain connectome with spatio-temporal attention.
\newblock {\em Advances in Neural Information Processing Systems}, 34:4314--4327, 2021.

\bibitem{kliemann2019intrinsic}
Dorit Kliemann, Ralph Adolphs, J~Michael Tyszka, Bruce Fischl, BT~Thomas Yeo, Remya Nair, Julien Dubois, and Lynn~K Paul.
\newblock Intrinsic functional connectivity of the brain in adults with a single cerebral hemisphere.
\newblock {\em Cell reports}, 29(8):2398--2407, 2019.

\bibitem{li2019global}
Jingwei Li, Ru~Kong, Rapha{\"e}l Li{\'e}geois, Csaba Orban, Yanrui Tan, Nanbo Sun, Avram~J Holmes, Mert~R Sabuncu, Tian Ge, and BT~Thomas Yeo.
\newblock Global signal regression strengthens association between resting-state functional connectivity and behavior.
\newblock {\em NeuroImage}, 196:126--141, 2019.

\bibitem{li2019graph}
Xiaoxiao Li, Nicha~C Dvornek, Yuan Zhou, Juntang Zhuang, Pamela Ventola, and James~S Duncan.
\newblock Graph neural network for interpreting task-fmri biomarkers.
\newblock In {\em Medical Image Computing and Computer Assisted Intervention--MICCAI 2019: 22nd International Conference, Shenzhen, China, October 13--17, 2019, Proceedings, Part V 22}, pages 485--493. Springer, 2019.

\bibitem{li2021braingnn}
Xiaoxiao Li, Yuan Zhou, Nicha Dvornek, Muhan Zhang, Siyuan Gao, Juntang Zhuang, Dustin Scheinost, Lawrence~H Staib, Pamela Ventola, and James~S Duncan.
\newblock Braingnn: Interpretable brain graph neural network for fmri analysis.
\newblock {\em Medical Image Analysis}, 74:102233, 2021.

\bibitem{li2020pooling}
Xiaoxiao Li, Yuan Zhou, Nicha~C Dvornek, Muhan Zhang, Juntang Zhuang, Pamela Ventola, and James~S Duncan.
\newblock Pooling regularized graph neural network for fmri biomarker analysis.
\newblock In {\em Medical Image Computing and Computer Assisted Intervention--MICCAI 2020: 23rd International Conference, Lima, Peru, October 4--8, 2020, Proceedings, Part VII 23}, pages 625--635. Springer, 2020.

\bibitem{liu2023braintgl}
Lingwen Liu, Guangqi Wen, Peng Cao, Tianshun Hong, Jinzhu Yang, Xizhe Zhang, and Osmar~R Zaiane.
\newblock Braintgl: A dynamic graph representation learning model for brain network analysis.
\newblock {\em Computers in Biology and Medicine}, 153:106521, 2023.

\bibitem{liu2018subcortical}
Xiao Liu, Jacco~A De~Zwart, Marieke~L Sch{\"o}lvinck, Catie Chang, Frank~Q Ye, David~A Leopold, and Jeff~H Duyn.
\newblock Subcortical evidence for a contribution of arousal to fmri studies of brain activity.
\newblock {\em Nature communications}, 9(1):395, 2018.

\bibitem{makovac2019response}
Elena Makovac, Sabrina Fagioli, David~R Watson, Frances Meeten, Jonathan Smallwood, Hugo~D Critchley, and Cristina Ottaviani.
\newblock Response time as a proxy of ongoing mental state: A combined fmri and pupillometry study in generalized anxiety disorder.
\newblock {\em Neuroimage}, 191:380--391, 2019.

\bibitem{margulies2016situating}
Daniel~S Margulies, Satrajit~S Ghosh, Alexandros Goulas, Marcel Falkiewicz, Julia~M Huntenburg, Georg Langs, Gleb Bezgin, Simon~B Eickhoff, F~Xavier Castellanos, Michael Petrides, et~al.
\newblock Situating the default-mode network along a principal gradient of macroscale cortical organization.
\newblock {\em Proceedings of the National Academy of Sciences}, 113(44):12574--12579, 2016.

\bibitem{parkes2018evaluation}
Linden Parkes, Ben Fulcher, Murat Y{\"u}cel, and Alex Fornito.
\newblock An evaluation of the efficacy, reliability, and sensitivity of motion correction strategies for resting-state functional mri.
\newblock {\em Neuroimage}, 171:415--436, 2018.

\bibitem{power2011functional}
Jonathan~D Power, Alexander~L Cohen, Steven~M Nelson, Gagan~S Wig, Kelly~Anne Barnes, Jessica~A Church, Alecia~C Vogel, Timothy~O Laumann, Fran~M Miezin, Bradley~L Schlaggar, et~al.
\newblock Functional network organization of the human brain.
\newblock {\em Neuron}, 72(4):665--678, 2011.

\bibitem{preti2017dynamic}
Maria~Giulia Preti, Thomas~AW Bolton, and Dimitri Van De~Ville.
\newblock The dynamic functional connectome: State-of-the-art and perspectives.
\newblock {\em Neuroimage}, 160:41--54, 2017.

\bibitem{radford2018improving}
Alec Radford, Karthik Narasimhan, Tim Salimans, Ilya Sutskever, et~al.
\newblock Improving language understanding by generative pre-training.
\newblock 2018.

\bibitem{radford2019language}
Alec Radford, Jeffrey Wu, Rewon Child, David Luan, Dario Amodei, Ilya Sutskever, et~al.
\newblock Language models are unsupervised multitask learners.
\newblock {\em OpenAI blog}, 1(8):9, 2019.

\bibitem{raut2021global}
Ryan~V Raut, Abraham~Z Snyder, Anish Mitra, Dov Yellin, Naotaka Fujii, Rafael Malach, and Marcus~E Raichle.
\newblock Global waves synchronize the brain’s functional systems with fluctuating arousal.
\newblock {\em Science advances}, 7(30):eabf2709, 2021.

\bibitem{said2023neurograph}
Anwar Said, Roza Bayrak, Tyler Derr, Mudassir Shabbir, Daniel Moyer, Catie Chang, and Xenofon Koutsoukos.
\newblock Neurograph: Benchmarks for graph machine learning in brain connectomics.
\newblock {\em Advances in Neural Information Processing Systems}, 36:6509--6531, 2023.

\bibitem{savva2019assessment}
Antonis~D Savva, Georgios~D Mitsis, and George~K Matsopoulos.
\newblock Assessment of dynamic functional connectivity in resting-state fmri using the sliding window technique.
\newblock {\em Brain and behavior}, 9(4):e01255, 2019.

\bibitem{schaefer2018local}
Alexander Schaefer, Ru~Kong, Evan~M Gordon, Timothy~O Laumann, Xi-Nian Zuo, Avram~J Holmes, Simon~B Eickhoff, and BT~Thomas Yeo.
\newblock Local-global parcellation of the human cerebral cortex from intrinsic functional connectivity mri.
\newblock {\em Cerebral cortex}, 28(9):3095--3114, 2018.

\bibitem{shine2016dynamics}
James~M Shine, Patrick~G Bissett, Peter~T Bell, Oluwasanmi Koyejo, Joshua~H Balsters, Krzysztof~J Gorgolewski, Craig~A Moodie, and Russell~A Poldrack.
\newblock The dynamics of functional brain networks: integrated network states during cognitive task performance.
\newblock {\em Neuron}, 92(2):544--554, 2016.

\bibitem{smith2015positive}
Stephen~M Smith, Thomas~E Nichols, Diego Vidaurre, Anderson~M Winkler, Timothy~EJ Behrens, Matthew~F Glasser, Kamil Ugurbil, Deanna~M Barch, David~C Van~Essen, and Karla~L Miller.
\newblock A positive-negative mode of population covariation links brain connectivity, demographics and behavior.
\newblock {\em Nature neuroscience}, 18(11):1565--1567, 2015.

\bibitem{sporns2016modular}
Olaf Sporns and Richard~F Betzel.
\newblock Modular brain networks.
\newblock {\em Annual review of psychology}, 67:613--640, 2016.

\bibitem{thomas2022self}
Armin Thomas, Christopher R{\'e}, and Russell Poldrack.
\newblock Self-supervised learning of brain dynamics from broad neuroimaging data.
\newblock {\em Advances in Neural Information Processing Systems}, 35:21255--21269, 2022.

\bibitem{van2008visualizing}
Laurens Van~der Maaten and Geoffrey Hinton.
\newblock Visualizing data using t-sne.
\newblock {\em Journal of machine learning research}, 9(11), 2008.

\bibitem{van2012influence}
Koene~RA Van~Dijk, Mert~R Sabuncu, and Randy~L Buckner.
\newblock The influence of head motion on intrinsic functional connectivity mri.
\newblock {\em Neuroimage}, 59(1):431--438, 2012.

\bibitem{van2013wu}
David~C Van~Essen, Stephen~M Smith, Deanna~M Barch, Timothy~EJ Behrens, Essa Yacoub, Kamil Ugurbil, Wu-Minn~HCP Consortium, et~al.
\newblock The wu-minn human connectome project: an overview.
\newblock {\em Neuroimage}, 80:62--79, 2013.

\bibitem{vaswani2017attention}
Ashish Vaswani, Noam Shazeer, Niki Parmar, Jakob Uszkoreit, Llion Jones, Aidan~N Gomez, {\L}ukasz Kaiser, and Illia Polosukhin.
\newblock Attention is all you need.
\newblock {\em Advances in neural information processing systems}, 30, 2017.

\bibitem{vidaurre2019stable}
Diego Vidaurre, Mark~W Woolrich, Anderson~M Winkler, Theodoros Karapanagiotidis, Jonathan Smallwood, and Thomas~E Nichols.
\newblock Stable between-subject statistical inference from unstable within-subject functional connectivity estimates.
\newblock {\em Human brain mapping}, 40(4):1234--1243, 2019.

\bibitem{wang2007altered}
Kun Wang, Meng Liang, Liang Wang, Lixia Tian, Xinqing Zhang, Kuncheng Li, and Tianzi Jiang.
\newblock Altered functional connectivity in early alzheimer's disease: A resting-state fmri study.
\newblock {\em Human brain mapping}, 28(10):967--978, 2007.

\bibitem{wang2022bevt}
Rui Wang, Dongdong Chen, Zuxuan Wu, Yinpeng Chen, Xiyang Dai, Mengchen Liu, Yu-Gang Jiang, Luowei Zhou, and Lu~Yuan.
\newblock Bevt: Bert pretraining of video transformers.
\newblock In {\em Proceedings of the IEEE/CVF conference on computer vision and pattern recognition}, pages 14733--14743, 2022.

\bibitem{wig2017segregated}
Gagan~S Wig.
\newblock Segregated systems of human brain networks.
\newblock {\em Trends in cognitive sciences}, 21(12):981--996, 2017.

\bibitem{yan2019groupinn}
Yujun Yan, Jiong Zhu, Marlena Duda, Eric Solarz, Chandra Sripada, and Danai Koutra.
\newblock Groupinn: Grouping-based interpretable neural network for classification of limited, noisy brain data.
\newblock In {\em Proceedings of the 25th ACM SIGKDD international conference on knowledge discovery \& data mining}, pages 772--782, 2019.

\bibitem{yang2022data}
Yi~Yang, Yanqiao Zhu, Hejie Cui, Xuan Kan, Lifang He, Ying Guo, and Carl Yang.
\newblock Data-efficient brain connectome analysis via multi-task meta-learning.
\newblock In {\em Proceedings of the 28th ACM SIGKDD Conference on Knowledge Discovery and Data Mining}, pages 4743--4751, 2022.

\bibitem{yeo2011organization}
BT~Thomas Yeo, Fenna~M Krienen, Jorge Sepulcre, Mert~R Sabuncu, Danial Lashkari, Marisa Hollinshead, Joshua~L Roffman, Jordan~W Smoller, Lilla Z{\"o}llei, Jonathan~R Polimeni, et~al.
\newblock The organization of the human cerebral cortex estimated by intrinsic functional connectivity.
\newblock {\em Journal of neurophysiology}, 2011.

\bibitem{zhao2022dynamic}
Kanhao Zhao, Boris Duka, Hua Xie, Desmond~J Oathes, Vince Calhoun, and Yu~Zhang.
\newblock A dynamic graph convolutional neural network framework reveals new insights into connectome dysfunctions in adhd.
\newblock {\em Neuroimage}, 246:118774, 2022.

\end{thebibliography}

\clearpage
\appendix
\begin{figure*}[h]
\begin{center}
\includegraphics[width=1\textwidth]{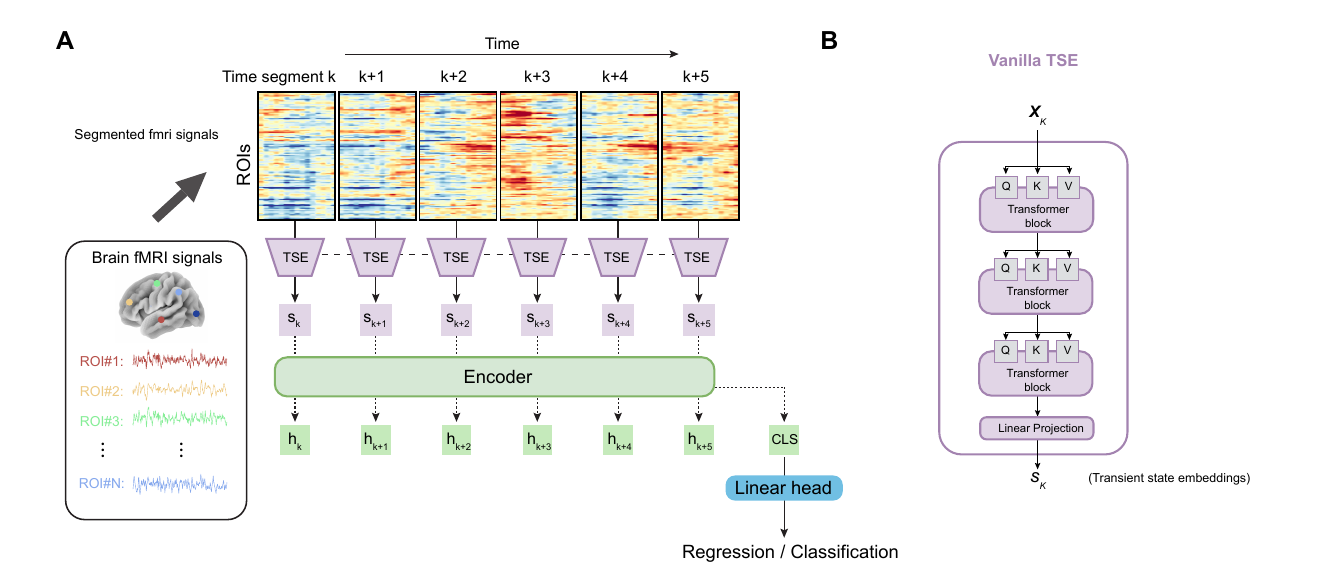}
\end{center}
\caption{A. BrainMAE fine-tuning framework. B. vanilla-TSE modules utilizing only the self-attention in the transformer blocks.}
\label{appdx-fig-5}
\end{figure*}

\section{Experiment Details}
\subsection{Pretraining}
\label{appdx-pretrain-setting}
Table~\ref{appdx-tbl-pretrain} provides a summary of our pretraining configurations, which are used for training BrainMAE on various datasets. Our ROIs are extracted based on Schaefer2018\_100ROIs Parcels, which include 100 cortical ROIs \cite{schaefer2018local}. During the training process, we employ a random selection method to choose a continuous segment of fMRI signals lasting for 300 seconds. For instance, we randomly select 300 consecutive fMRI signals from the original 864 seconds of data. This ensures that the signal used for masking and subsequent reconstruction is of equal size, allowing for the construction of mini-batches for parallelized training.

Additionally, this approach offers the advantage of efficient GPU memory utilization and scalability. It also introduces a degree of randomness, which, to some extent, serves as data augmentation and benefits representation learning.

\subsection{Transfer Learning}
\label{appdx-transfer-setting}
We fine-tune all BrainMAE models following the parameters outlined in Table~\ref{appdx-tbl-transfer}.  During both the training and testing phases, we utilize the original length of the fMRI data. For example, in the case of HCP3T with 864 seconds of data, we use the first 855 seconds, dividing it into 57 time segments that are fed into the model. The decoder is omitted during transfer learning. A task-specific linear head is appended to the CLS token representation generated by the encoder transformer for task-specific predictions, as shown in Figure~\ref{appdx-fig-5}A.

\begin{table}[t]
  \parbox{.5\linewidth}{
  \centering
  \caption{Pretraining Settings}
  \label{appdx-tbl-pretrain}
  \resizebox{0.95\linewidth}{!}{
    \begin{tabular}{ll}
    \toprule
    \addlinespace
    config	& value\\
    \addlinespace
    \hline
    \addlinespace
    optimizer & AdamW\\
    Training epochs	& 1000\\
    weight decay	& 0.05\\
    optim momentum	& $\beta1$, $\beta2$ = 0.9, 0.95\\
    Base learning rate	& 0.001\\
    $\lambda$	& 0.75\\
    batch size	& 32\\
    batch accumulation	& 4\\
    learning rate schedule	& cosine decay\\
    warmup epochs	& 100\\
    \bottomrule
    \end{tabular}
    }}
\hfill
    \parbox{.51\linewidth}{
  \centering
  \caption{Fine-tuning settings}
  \label{appdx-tbl-transfer}
  \resizebox{1\linewidth}{!}{
    \begin{tabular}{ll}
    \toprule
    \addlinespace
    config	& value\\
    \addlinespace
    \hline
    \addlinespace
    optimizer	& AdamW\\
    Training epochs	& 150\\
    Train:Val:Test (each fold)	& 0.64:0.2:0.16\\
    weight decay	& 0.05\\
    optim momentum	& $\beta1$, $\beta2$ = 0.9, 0.95\\
    Base learning rate	& 0.001\\
    batch size	& 64\\
    batch accumulation	& 2\\
    learning rate schedule	& cosine decay\\
    text{clip\_grad}	& 5\\
    \bottomrule
    \end{tabular}
}}
\end{table}

\section{Model Variants}
\subsection{vanilla-BrainMAE}
\label{appdx-fdgmae}
The vanilla-BrainMAE shares the exact same architecture as BrainMAE with the only exception being the use of vanilla-TSE to extract transient state embeddings from the fMRI segments. The vanilla-TSE, as shown in Figure~\ref{appdx-fig-5}, incorporates three standard transformer blocks that exclusively utilize self-attention. The vanilla-BrainMAE serves as a model for comparison with both AG-BrainMAE and SG-BrainMAE, allowing us to evaluate the proposed embedding-based static graph module.

\subsection{vanilla-BrainAE}
\label{appdx-fdgae}
The vanilla-BrainAE employs the exact same architecture as vanilla-BrainMAE, with the only distinction being the use of traditional autoencoding for fMRI signal reconstruction without signal masking. vanilla-BrainAE is included as a comparative model to assess the proposed masked autoencoding approach in comparison to the other models.

\begin{table}[h]
\setlength{\tabcolsep}{15 pt}
\renewcommand{\arraystretch}{1.1}
\caption{The distinction between different model variants}
\label{appdx-tbl-distinction-model-variants}
\resizebox{1\linewidth}{!}{
\centering
{\begin{tabularx}{\textwidth}{p{32mm}p{60mm}p{20mm}}
\toprule
Model variants&Transient state encoder (TSE)&Pretraining
\\ \hline 
\textit{\textbf {Primary}}\\
SG-BrainMAE & Three region-aware graph attention block & Masked-Autoencoding   \\
\addlinespace
AG-BrainMAE & Two self-attention blocks stacked on top of a region-aware graph attention attention block & Masked-Autoencoding   \\
\addlinespace
\hline 
\textit{\textbf {Other}}\\
vanilla-BrainMAE & Three self-attention blocks & Masked-Autoencoding   \\
\addlinespace
vanilla-BrainAE & Three self-attention blocks & Autoencoding   \\
\addlinespace
PosSG-BrainMAE & Three absolute position embedding informed attention block (see Appendix~\ref{appdx-ablation-embedding}) & Masked-Autoencoding   \\
\addlinespace
SG-BrainMAE (SL) & Three region-aware graph attention block where attentions are computed without self-loop removal (see Appendix~\ref{appdx-ablation-sl-removal}) & Masked-Autoencoding   \\
\bottomrule
\end{tabularx}}
}
\end{table}

\section{Datasets}
\begin{table}[h]
\centering
\caption{Dataset Statistics}
\label{appdx-tbl-dataset-stats}
\begin{tabular}{ccccc}
\toprule
\addlinespace
{} & HCP-3T & HCP-7T & HCP-Aging & NSD	\\
\addlinespace
\hline
\addlinespace
Number of subjects	& 897 & 184	& 725 & 8 \\
Number of sessions	& 3422 & 720 & 2400	& 3120 \\
Number of TRs & 1200 & 1200 & 478 & 301 \\
Orignal TR(s) & 0.72 & 1.00 & 0.80 & 1.00 \\
Number of TR interpolate to 1s & 864 & {} & 382	& {} \\
Type & Resting-state & Resting-state & Resting-state & TASK \\
\addlinespace	
\bottomrule
\end{tabular}
\end{table}

\label{appdx-dataset}
\textbf{HCP-3T/HCP-7T Datasets.}

The Human Connectome Project (HCP) is a freely shared dataset from 1200 young adult (ages 22-35) subjects, using a protocol that includes structural images (T1w and T2w), functional magnetic resonance imaging (resting-state fMRI, task fMRI), and high angular resolution diffusion imaging (dMRI) at 3 Tesla (3T) and behavioral and genetic testing. Moreover, 184 subjects also have 7T MR scan data available (in addition to 3T MR scans),  which includes resting-state fMRI, retinotopy fMRI, movie-watching fMRI, and dMRI. In our study, we focused on both the resting-state and task sessions of the dataset as well as 8 behavior measurements (see Table \ref{appdx-tbl-beh} for more information).

\textbf{HCP-Aging Dataset.}

The Human Connectome Project Aging (HCP-Aging) dataset is an extensive and longitudinally designed neuroimaging resource aimed at comprehensively investigating the aging process within the human brain.  It comprises a wide array of multimodal neuroimaging data, such as structural MRI (sMRI), resting-state functional MRI (rs-fMRI), task-based fMRI (tfMRI), and diffusion MRI (dMRI), alongside rich cognitive and behavioral assessments. In our study, we focus on the resting-state session of the dataset as well as age and gender information.

\textbf{NSD Dataset.}

The Natural Scenes Dataset comprises whole-brain 7T functional magnetic resonance imaging (fMRI) scans at a high resolution of 1.8 mm. These scans were conducted on eight meticulously selected human participants, each of whom observed between 9,000 and 10,000 colorful natural scenes, totaling 22,000 to 30,000 trials, over the span of one year. While viewing these images, subjects were engaged in a continuous recognition task in which they reported whether they had seen each given image at any point in the experiment. 

Table~\ref{appdx-tbl-dataset-stats} provides an overview of the statistical information for each of the datasets employed in our study.

\section{ROI Embedding Analysis}

\label{appdx-emb-const}
\subsection{Relationship to Principal Gradient}
\begin{figure}[h]
\begin{center}
\includegraphics[width=1\textwidth]{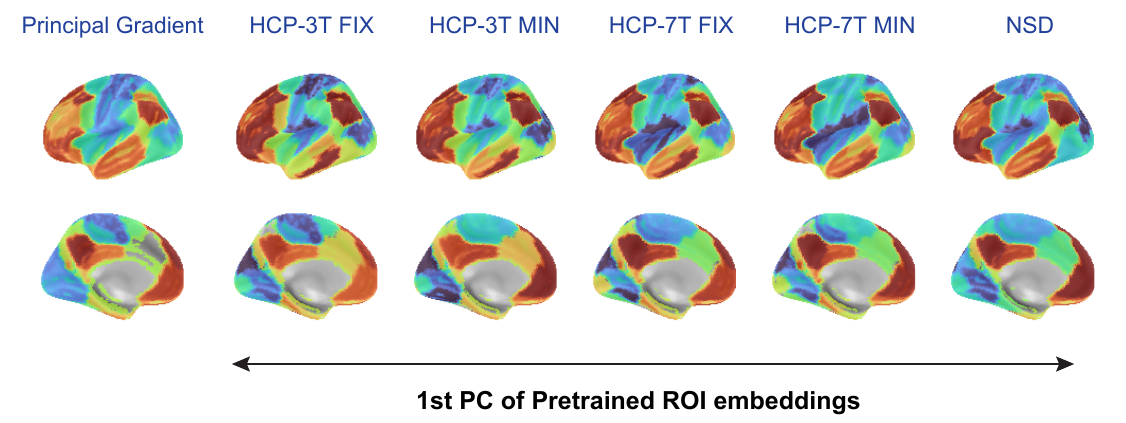}
\end{center}
\caption{Comparison between principal gradient and the first principal component of pretrained ROI embeddings on different datasets. The color mapped on the brain surface encodes either principal gradient value (1st column) or 1st PC value of pretrain embeddings (2-6th columns).}
\label{appdx-fig-6}
\end{figure}

\begin{figure}[h]
\begin{center}
\includegraphics[width=0.8\linewidth]{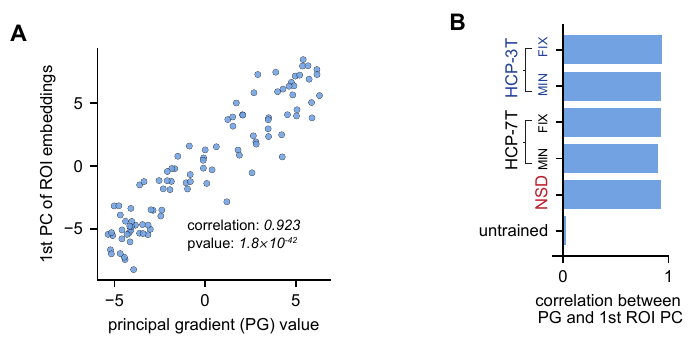}
\end{center}
\caption{A. Strong correlation between 1st PC of ROI embedding (HCP-3T pretrained) and principal gradient. B. This relationship is highly reproducible across differently pretrained models but is not the case for untrained model.}
\label{appdx-fig-7}
\end{figure}

In neuroscience research, the principal gradient characterize the topographical organization of brain regions, reflecting the between network functional organization \cite{margulies2016situating}. Along this principal direction, one end is associated with regions serving primary sensory/motor functions, while the other end corresponds to transmodal regions, often referred to as the default-mode network (DMN).

We have identified a significant relationship between the first principal component of pretrained ROI embeddings and the principal gradient. This finding suggests that the functional connectivity between brain networks is inherently encoded in these embeddings. Furthermore, this relationship is highly reproducible across pretrained models trained on different datasets, as shown in Figure~\ref{appdx-fig-6} and~\ref{appdx-fig-7}.
\subsection{Consistency Across Pretrained Models}

\begin{figure}[h]
\begin{center}
\includegraphics[width=0.7\linewidth]{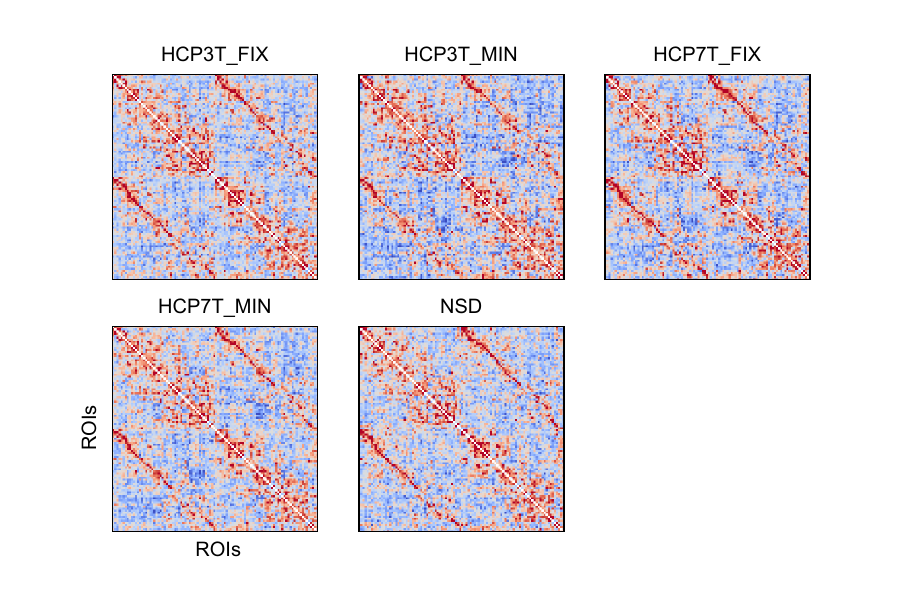}
\end{center}
\caption{The convergence result observed in the cross-regional embedding similarity matrix reveals the consistency of ROI embeddings pretrained on different datasets.}
\label{appdx-fig-8}
\end{figure}

We analyze the cross-region embedding similarity, or embeddings similarity matrix for each of the models pretrained on various datasets. We use the cosine distance to measure the similarity between two ROI embedding vectors. As shown in Figure~\ref{appdx-fig-8}, The embedding similarity matrices shows converging results on differently pretrained models, suggesting that highly similar embedding profiles can be identified in different datasets, thereby validating our hypothesis and the proposed approach.

\subsection{Age Effects}

\begin{figure}
\begin{center}
\includegraphics[width=0.6\linewidth]{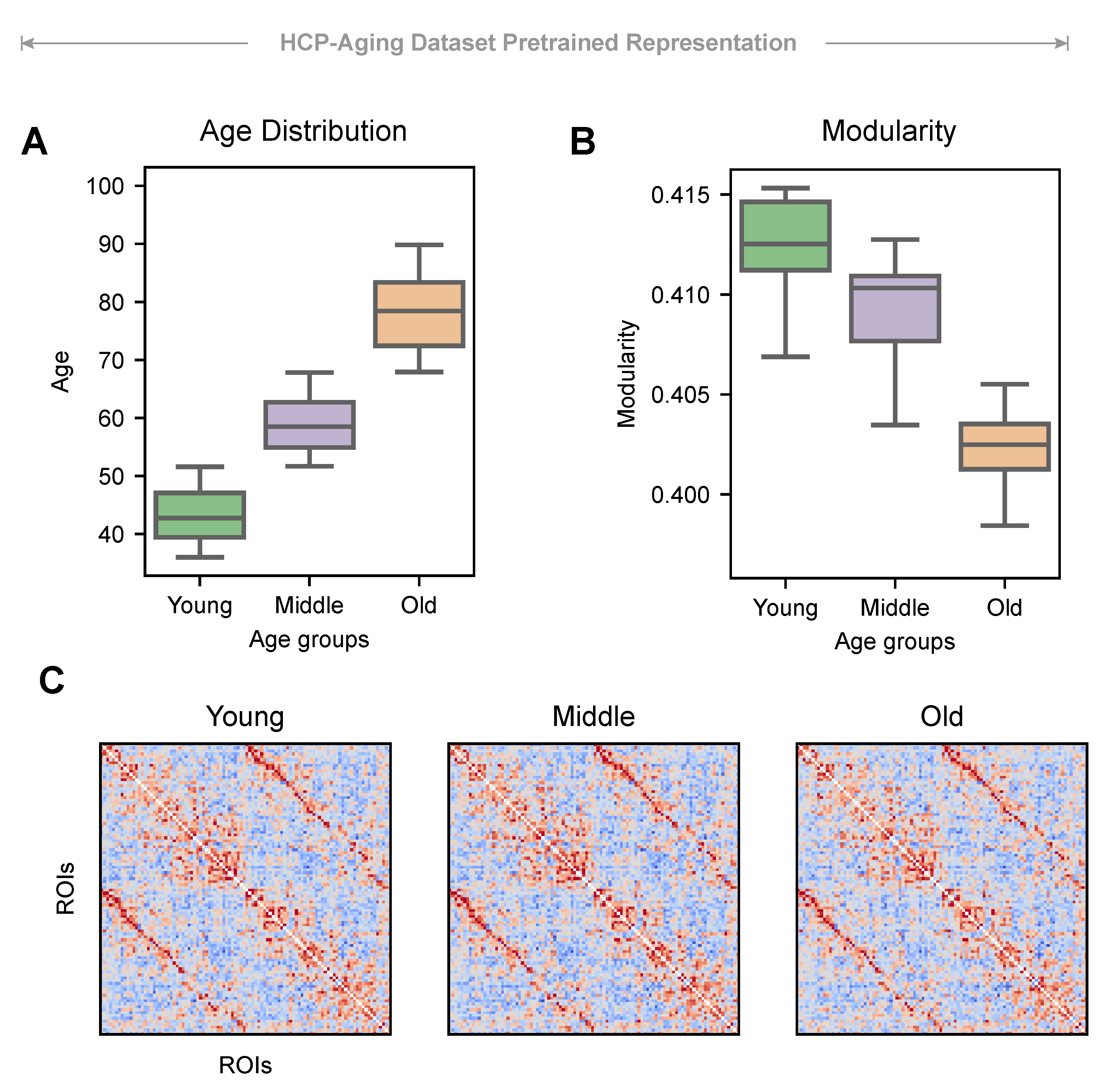}
\end{center}
\caption{Age effect on learned ROI embeddings. (A) Age distribution for the three defined age groups. (B) The modularity of the ROI embedding-based functional network decreases with aging. For each group, modularity is computed 50 times using the Louvain algorithm, and the whiskers of the boxplot represent the maximum and minimum of the 50 modularity values. (C)  Embedding similarity matrix of the three age groups.}
\label{appdx-fig-9}
\end{figure}

\label{appdx-age-effect}
The ROI embedding faithfully captures the characteristics of brain ROIs within the pre-trained fMRI dataset. To investigate the impact of aging on the acquired ROI embeddings, we partition the HCP-Aging dataset into three distinct, non-overlapping age groups (Young: 36-52, Middle: 52-68, Old: 68-100; refer to the Figure~\ref{appdx-fig-9}A for the age distribution in each group). Subsequently, we independently pre-train the SG-BrainMAE for each age group. To discern variations in ROI embeddings, we study the modular structure of the network constructed based on the embedding similarity matrix specific to each age group. The modularity of a network, serving as a metric in network analysis, delineates how the network can be segmented into nonoverlapping regions or the segregation of brain ROIs. The results in Figure~\ref{appdx-fig-9}B indicate a reduction in the modularity of ROI embeddings from the young age group to the old age group. This trend suggests a decline in the segregation of functional networks during aging, aligning with established findings in the neuroscience literature \cite{sporns2016modular, wig2017segregated}.

\section{fMRI Representation Analysis}
\subsection{Brain States}
\label{appdx-brain-state}
\subsubsection{Task vs. Rest}
\begin{figure*}[h]
\begin{center}
\includegraphics[width=1\textwidth]{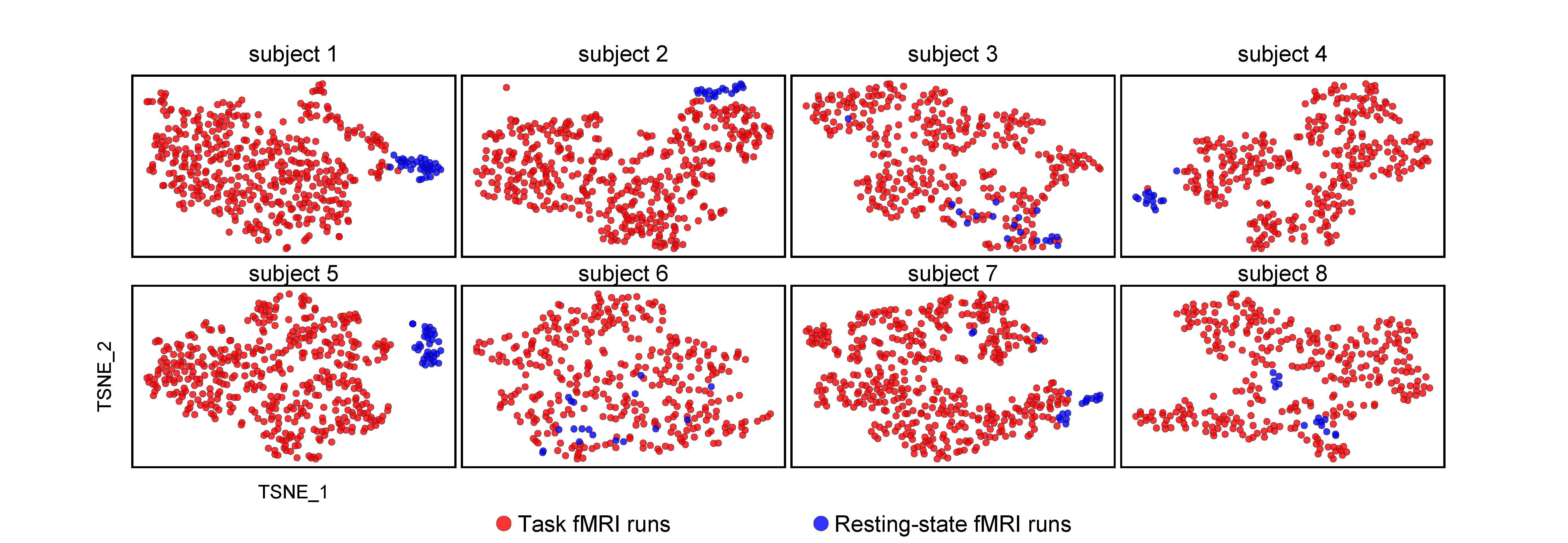}
\end{center}
\caption{Within-subject t-SNE plot using the pretrained SG-BrainMAE on the independent NSD dataset, with red indicating task-related fMRI runs and blue representing resting-state runs.}
\label{appdx-fig-10}
\end{figure*}
Taking a step further, we aimed to analyze the representation of fMRI scans within each subject. In the NSD dataset, each subject performed multiple task sessions (image-memory task) as well as a few resting-state sessions (spontaneous rest). The Figure~\ref{appdx-fig-10}B shows the within-subject t-SNE analysis of the representation extracted by the SG-MAE pretrained with HCP3T. Remarkably, in most cases, the resting-state fMRI runs are well separated from task fMRI runs and exhibit distinct representations. This result further suggests that within individuals, the pre-trained representation carries meaningful information reflecting one's brain state.

\subsubsection{Brain Arousal State}
\begin{figure*}[h]
\begin{center}
\includegraphics[width=1\textwidth]{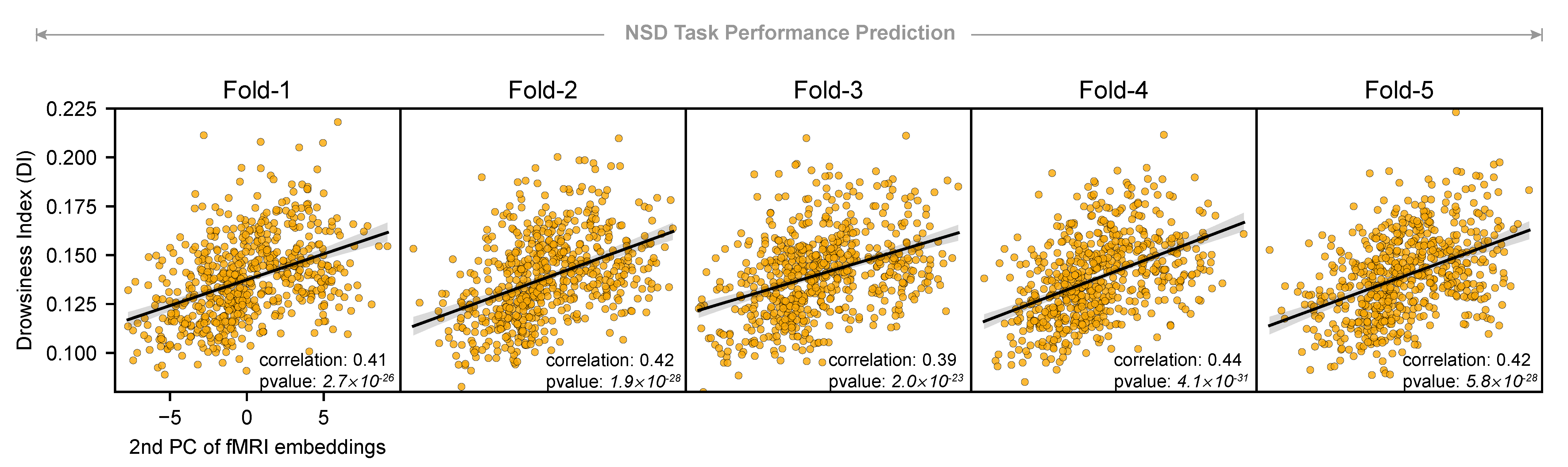}
\end{center}
\caption{Strong correlation between the second principal component (PC) of the learned fMRI representation and drowsiness index (DI). Each column displays the result of each fold in the 5-fold cross-validation.}
\label{appdx-fig-11}
\end{figure*}
In interpreting the representations extracted by the model fine-tuned on the NSD dataset for downstream task performance prediction, we conducted PCA analysis on the fMRI representations (output of CLS token). Intriguingly, as shown in Figure~\ref{appdx-fig-11}, we observed a close relationship between the second principal component and the drowsiness index, a metric for measuring the brain arousal level \cite{chang2016tracking, gu2020transient}. This finding suggests a convergence between our data-driven approach and traditional neuroscience studies that quantify brain states with multimodal equipment. It implies that the proposed method could serve as a valuable tool to sufficiently identify brain states using fMRI alone, obviating the need for additional modalities such as EEG.

\section{HCP Transient Mental State Decoding}
Table~\ref{appdx-tbl-mental-state-hcptask-detial} presents detailed results for each mental state from the multi-class decoding task, demonstrating consistently high decoding accuracy that appears to be insensitive to variations in state duration.

\begin{table*}[h]

\setlength{\tabcolsep}{4 pt}
\renewcommand{\arraystretch}{1.2}
\caption{Results for transient mental state decoding.}
\label{appdx-tbl-mental-state-hcptask-detial}
\resizebox{1\linewidth}{!}{
\centering
{\begin{tabular}{lllccccc}
\toprule
\multirow{2.5}{*}{Task}&\multicolumn{2}{c}{Mental States} && \multicolumn{4}{c}{Model F1-score(\%)}\\
\cmidrule(lr){2-3}\cmidrule(lr){5-8} 
&{Name}&{Duration(second)}&{}&{SG-BrainMAE}&{AG-BrainMAE}&{vanilla-BrainMAE}&{vanilla-BrainAE} \
\\ \hline 

\multirow{4}{*}{Working Memory} & body & 27.5 &{}&94.98±1.78 & 95.02±1.19 & 93.58±3.19 & 89.42±1.88\\
                                &faces & 27.5 &{}&95.80±2.28 & 96.11±1.65 & 95.52±0.99 & 92.50±1.98\\
                                &places &27.5 &{}& 97.47±1.25 & 98.20±1.75 & 96.94±1.59 & 96.12±2.72\\
                                &tools & 27.5&{}& 94.84±2.62 & 95.15±1.33 & 93.19±2.85 & 88.45±4.44\\
\hline 
\multirow{2}{*}{Gambling} & win & 28.0 &{}& 86.86±2.40 & 89.63±3.16 & 86.52±2.46 & 82.07±3.04\\
                        & loss & 28.0 & {}&87.27±1.09 & 89.21±2.88 & 87.57±1.98 & 80.90±3.77\\
\hline 
\multirow{5}{*}{Motor} & left finger & 12.0 & {}&98.95±1.46 & 99.25±1.31 & 98.67±1.23 & 98.40±1.71\\
                                &right finger & 12.0 & {}&98.03±1.77 & 98.04±1.76 & 97.19±2.08 & 97.45±2.29\\
                                &left toe &12.0 & {}&98.20±1.58 & 98.35±1.63 & 96.97±2.46 & 97.16±2.58\\
                                &right toe &12.0 & {}&97.62±2.06 & 98.21±1.87 & 96.99±2.56 & 97.08±2.71\\
                                &tongue &12.0 & {}&98.81±1.87 & 98.66±1.45 & 97.96±2.03 & 98.54±1.72\\
\hline 
\multirow{2}{*}{Language} & story & 25.9 & {}&97.52±1.93 & 97.73±2.14 & 97.58±1.68 & 97.73±1.24\\
                                &math & 16.0 & {}&98.86±0.99 & 98.96±0.93 & 98.65±0.56 & 98.57±0.67\\
\hline 
\multirow{2}{*}{Social} & interaction &23.0 & {}&98.22±1.38 & 97.84±1.02 & 97.95±0.93 & 97.59±0.88\\
                                & no interaction &23.0 & {}&98.55±1.38 & 97.61±1.57 & 97.64±1.38 & 96.92±1.14\\
\hline 
\multirow{2}{*}{Relational} & relational & 16.0& {}&93.74±1.89 & 94.20±2.43 & 92.48±3.47 & 90.83±3.45\\
                                &matching &16.0 & {}&93.83±2.25 & 93.42±1.62 & 92.03±2.32 & 90.58±2.02\\
\hline 
\multirow{2}{*}{Emotion} & fear & 18.0&{}& 97.08±1.35 & 97.69±1.71 & 96.20±2.15 & 96.51±1.77\\
                                & neutral & 18.0 & {}& 97.39±1.78 & 97.89±1.41 & 97.74±1.20 & 97.24±1.23\\
\hline 
Rest & Rest & 864.0 & {} & 80.82±9.76 & 80.62±1.05 & 87.24±5.46 & 82.05±6.12\\
\bottomrule
\end{tabular}}
}
\end{table*}

\section{Simulation Analysis}
\label{appdx-simulation-analysis}
We first define N networks (or communities), where nodes within each network are functionally connected and exhibit similar fluctuations.  We then assign each Region of Interest (ROI) a vector that represents the probability of belonging to each network. For every synthetic fMRI run, ROIs are allocated to networks based on their probability (the higher the probability, the greater the likelihood of the ROI being assigned to that network). ROIs within the same network are assumed to exhibit identical fluctuation profiles, adhering to the principle of "wire together, fire together". We simulate the time series for each ROI using a time-varying sine wave. ROIs within the same network share a common phase with small random perturbations, whereas ROIs from different networks are characterized by distinct phase profiles. Across different synthetic fMRI runs, the same ROI may be classified into different networks.

For the experiment, we utilize N=10 networks and 100 ROIs. SG-BrainMAE is pre-trained on this synthetic dataset with hyper-parameters same to the setting described in Appendix~\ref{appdx-pretrain-setting} and Table~\ref{appdx-tbl-pretrain}.

\section{Additional Results}
\subsection{Ablation Study on ROI Embedding}
\label{appdx-ablation-embedding}
Given that the position of the ROIs is static, there exists a possibility that the learned ROI embedding may predominantly encode information about the absolute position rather than the functional characteristics of the ROI. This hypothesis is investigated with two analyses from different perspectives.

\begin{figure*}[h]
\begin{center}
\includegraphics[width=0.8\textwidth]{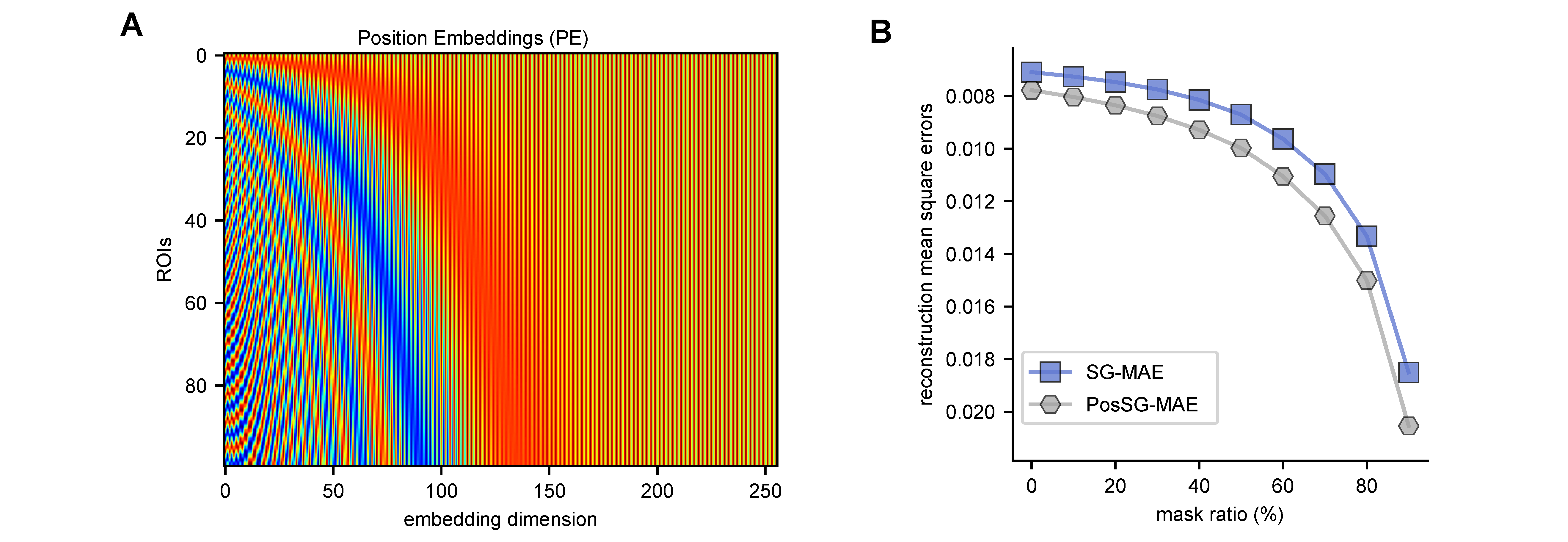}
\end{center}
\caption{Ablation analysis 1: replacing ROI embeddings with position embeddings. (A) Position embedding used as ROI embeddings.
(B) The PosSG-BrainMAE, utilizing position embeddings, exhibited inferior reconstruction results compared to SG-BrainMAE with informative ROI embeddings.}
\label{appdx-fig-8-recon}
\end{figure*}

\textbf{Analysis 1.} We substitute ROI embedding in the SG-BrainMAE model with position embedding, which is then frozen throughout the pretraining phase on the HCP-3T dataset while keeping other model components unchanged. This adapted model is named PosSG-BrainMAE. Evaluations involving both the reconstruction of masked signals on an independent HCP-7T dataset (see Figure~\ref{appdx-fig-8-recon}) and performance in downstream tasks (see Tabel~\ref{appdx-tbl-ablation-analysis-1-hcp3t}, \ref{appdx-tbl-ablation-analysis-1-hcpaging}, and \ref{appdx-tbl-ablation-analysis-1-nsd}) reveal a decreased efficacy compared to SG-BrainMAE. This decline in performance justifies the valuable ROI information contained in learned ROI embeddings, extending beyond mere positional information.

\begin{figure*}[h]
\begin{center}
\includegraphics[width=0.8\textwidth]{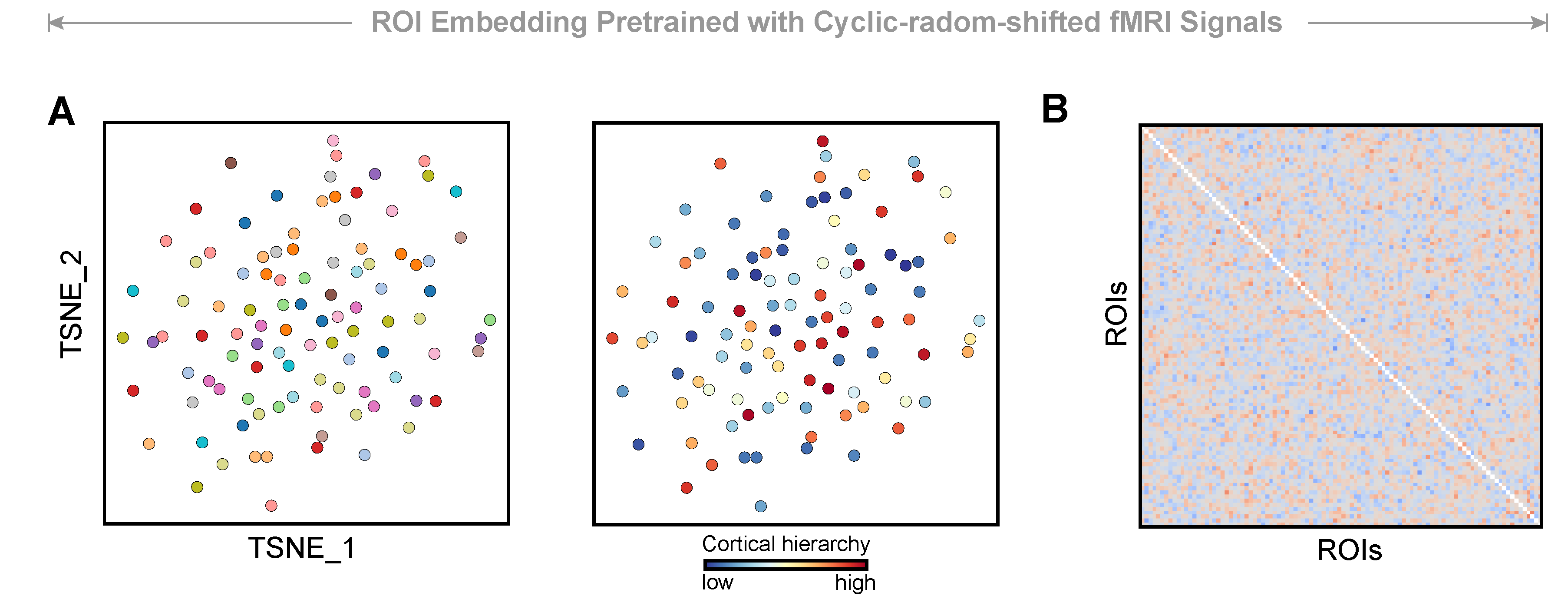}
\end{center}
\caption{Ablation analysis 2: evaluation of pre-trained ROI embeddings with cyclic-random-shifted fMRI signals. (A) A t-SNE plot of ROI embeddings with ROI colored based on Yeo-17 network (Left) and cortical hierarchy (Right). (B) ROI embedding similarity matrix with indiscernible inter-network relationship.}
\label{appdx-fig-12-cyclshift}
\end{figure*}
\textbf{Analysis 2.} We cyclic shift each ROI's fMRI signal by random time steps. By doing this, the shifted fMRI signal exhibit two properties: 1.  each fMRI signal itself is merely changed; 2. elimination of the inter-relationship between pairs of ROIs. We then follow the same pre-training procedure using this modified dataset. The ROI embeddings learned from this dataset did not exhibit functional specificity and inter-regional connectivity (see Figure~\ref{fig-3}), in contrast to those learned from the actual dataset (see Figure~\ref{appdx-fig-12-cyclshift}). These findings provide additional evidence that the proposed method learns meaningful ROI information from the dataset, including its relationship to other ROIs.

\subsection{Ablation Study on Self-Loop Removal}
\label{appdx-ablation-sl-removal}
Removing self-loops in our model avoids the attention mechanism favoring its own node, encouraging it to aggregate signals from other relevant nodes. This approach is akin to a group voting system and helps reduce sensitivity to input noise. To validate this design choice, we conducted a comparative analysis for downstream tasks between SG-BrainMAE and a similar model that includes self-loops, named SG-BrainMAE(SL). The results, presented in Tables~\ref{appdx-tbl-ablation-analysis-1-hcp3t}, \ref{appdx-tbl-ablation-analysis-1-hcpaging}, and \ref{appdx-tbl-ablation-analysis-1-nsd}, show a slight decrease in performance for SG-BrainMAE(SL), indicating the effectiveness of excluding self-loops in our model.

\begin{table*}[h]
\setlength{\tabcolsep}{4 pt}
\renewcommand{\arraystretch}{1.2}
\begin{center}
\caption{More ablation study: behavior prediction.}
\label{appdx-tbl-ablation-analysis-1-hcp3t}
\resizebox{1\linewidth}{!}{
{\begin{tabular}{llllllllllll}
\toprule
\multirow{2.5}{*}{Model}&\multicolumn{2}{c}{Gender}&&\multicolumn{8}{c}{Behaviors (measured in MAE)}\\
\cmidrule(lr){2-3} \cmidrule(lr){5-12} 
&{Accuracy(\%)}&{AUROC}&{}&{PicSeq}&{PMT$\_$CR}&{PMT$\_$SI}&{PicVocab}&{IWRD}&{ListSort}&{LifeSatisf}&{PSQI} \
\\ \hline 
\addlinespace
\rowcolor{Gray}
SG-BrainMAE & \bf 97.49±0.15 & \bf 97.46±0.18 &{}& \bf 5.06±0.21 & \bf 1.63±0.08 & \bf 1.24±0.04 & \bf 3.40±0.14 & \bf 1.11±0.04 & \bf 4.35±0.12 & \bf 3.64±0.27 & \bf 1.05±0.06\\
PosSG-BrainMAE & 95.74±1.08 & 95.98±0.96 &{}& 5.99±0.30 & 1.83±0.09 & 1.41±0.08 & 3.80±0.08 & 1.28±0.05  & 4.83±0.20 & 4.02±0.08 & 1.22±0.08 \\
SG-BrainMAE(SL) & 96.63±1.56 & 96.62±1.55 &{}& 5.53±0.15 & 1.72±0.06 & 1.32±0.07 & 3.61±0.10 & 1.15±0.03 & 4.64±0.06 & 3.87±0.10 & 1.14±0.04\\
\bottomrule
\end{tabular}}
}
\end{center}
\end{table*}

\begin{table}[h]
  \parbox{.5\linewidth}{
  \centering
  \caption{More ablation study: age prediction.}
  \label{appdx-tbl-ablation-analysis-1-hcpaging}
  \resizebox{1\linewidth}{!}{
    \begin{tabular}{lllll}
    \toprule
    \multirow{2.5}{*}{ Model}&\multicolumn{2}{c}{ Gender}&{}&\multirow{2.5}{*}{ Aging(MAE)}\\
    \cmidrule(lr){2-3} 
    &{Accuracy(\%)}&{AUROC}
    \\ \hline 
    \addlinespace
    \rowcolor{Gray}
    SG-BrainMAE& \bf 92.67±1.07 & \bf 92.51±1.07 &{}& \bf 5.75±0.44 \\
    PosSG-BrainMAE & 88.38±2.93 & 88.28±3.19 &{}& 6.66±0.71  \\
    SG-BrainMAE(SL) & 91.29±1.34 & 91.45±1.33 &{}&5.68±0.31\\
    \bottomrule
    \end{tabular}
    }}
    \hfill
    \parbox{.5\linewidth}{
    \centering
    \caption{Task performance prediction.}
    \label{appdx-tbl-ablation-analysis-1-nsd}
    \resizebox{0.8\linewidth}{!}{
    \begin{tabular}{lll}
    \toprule
    \multirow{2.5}{*}{Model}&\multirow{2.5}{*}{ Task Accuracy}&\multirow{2.5}{*}{ RT(ms)}\\
    \addlinespace
    \\ \hline 
    \addlinespace
    \rowcolor{Gray}
    SG-BrainMAE& \bf 0.069±0.004 & \bf 90.678±1.767  \\
    PosSG-BrainMAE & 0.080±0.004 & 100.064±4.439  \\ 
    SG-BrainMAE(SL) & 0.078±0.003 & 98.469±1.675\\
    \bottomrule
    \end{tabular}
    }}
\end{table}

\subsection{Comparison with Self-supervised Learning on Gender Classification on HCP-3T dataset}
We conducted comparative analysis between our method and another recent self-supervised learning approach, named TFF, which employs 3D Convolutional Neural Networks (CNNs) and transformer to extract volumetric representations of fMRI data at each time point, with pre-training via auto-encoding. The results, as shown in Table~\ref{appdx-tbl-tff}, demonstrate that our model outperforms TFF in the HCP gender prediction downstream task.

\begin{table}[!h]
\setlength{\tabcolsep}{8 pt}
\renewcommand{\arraystretch}{1.1}
\caption{HCP: gender classification}
\label{appdx-tbl-tff}
\centering
\resizebox{0.3\linewidth}{!}{
{\begin{tabular}{lll}
\toprule
Model&Accuracy(\%)
\\ \hline 
TFF & 94.09  \\
\hline 
\textit{\textbf {Ours}}\\
\rowcolor{Gray}
\bf SG-BrainMAE & \bf 97.49±0.15  \\
\rowcolor{Gray}
\bf AG-BrainMAE & \bf 97.13±0.56 \\
\bottomrule
\end{tabular}}
}
\end{table}

\subsection{Comparison with Traditional Machine Learning Models}
Given the effectiveness and prevalence of traditional machine learning (ML) models in neuroimaging communities, this section focuses on assessing the added performance benefits of utilizing complex deep learning-based methods in comparison to these simpler ML models. 

For regression tasks, we consider a suite of linear models, including ordinary linear regression, ridge regression, and elastic net. In the context of classification tasks, we explore the use of logistic regression, linear Support Vector Machine (SVM), and Random Forest models. Each of these models is trained to make predictions based on the flattened upper-triangle of the Functional Connectivity (FC) connectivity matrix.

We employ cross-validated grid search approach, with specific ranges and increments for each model for hyperparameter selection of each ML model:
\begin{itemize}
    \item \textbf{Support Vector Machine (SVM)}: We vary the L2 regularization coefficient from 0.1 to 10, with an increment of 0.5.

    \item \textbf{Logistic Regression}: The L2 regularization coefficient is tuned from 0.1 to 10, with an increment of 0.5.

    \item \textbf{Random Forest}: Three key parameters are tuned:
    a. Number of trees, ranging from 1 to 250 with increments of 50.
    b. Maximum depth of each tree, from 5 to 50 with increments of 10.
    c. Minimum samples required to split a node, from 5 to 100 with increments of 20.

    \item \textbf{Ordinary Linear Regression}: This model did not require hyperparameter tuning.

    \item \textbf{Ridge Regression}: The L2 regularization coefficient is tuned from 0 to 10, with an increment of 0.5.

    \item \textbf{Elastic Net Regression}:
a. The coefficient of the L2 penalty (ridge regression component) is tuned from 0 to 10, with increments of 0.5.
b. The coefficient of the L1 penalty (lasso regression component) is adjusted from 0 to 1, with increments of 0.2.
\end{itemize}
For models requiring multiple hyperparameters, we train for each possible combination. The best-performing model is selected based on its performance on the validation set, using Mean Squared Error (MSE) for regression tasks and accuracy for classification tasks.

The results of this comparative analysis across three different downstream tasks are shown in Table~\ref{appdx-tbl-all-hcp3t}, \ref{appdx-tbl-all-hcp3t-beh}, \ref{appdx-tbl-all-hcpaging} and \ref{appdx-tbl-all-nsd}. It reveals that for classification tasks, traditional ML methods demonstrate performance levels comparable to those of baseline deep learning methods. This observation can be attributed to the well-established understanding that the functional connectivity matrix harbors significant information pertinent to human traits and age. However, for more complex regression tasks, such as task performance prediction, which necessitate inferring intricate brain states from the dynamics of fMRI signals, ML models often exhibit less satisfactory performance. In such scenarios, deep learning methods, endowed with their robust capability for representation learning, are able to achieve markedly superior results.

\begin{table*}[h]
\setlength{\tabcolsep}{4 pt}
\renewcommand{\arraystretch}{1.2}
\centering
\caption{HCP-3T gender classification}
\label{appdx-tbl-all-hcp3t}
\resizebox{0.8\linewidth}{!}{
{\begin{tabular}{lllllll}
\toprule
\multirow{2.5}{*}{Model}&\multicolumn{5}{c}{Gender}\\
\cmidrule(lr){2-6} 
&{Accuracy(\%)}&{AUROC}&{specificity(\%)}&{sensitivity(\%)}&{F1 Score(\%)} \
\\ \hline 
\textit{\textbf {FIX-FC}}\\
BrainNetTF & 94.11±0.98 & 94.39±0.90 & 95.36±0.70 & 93.24±2.08 & 93.69±1.01\\
VanillaTF & 90.00±1.05 & 89.93±0.94 & 91.36±1.25 & 88.32±2.51 & 88.71±1.10 \\
BrainNetCNN & 90.68±1.80 & 90.89±1.55 & 93.82±1.64 & 87.40±3.80 & 89.56±1.44\\
 \hline 
\textit{\textbf {Dynamic-FC}}\\
STAGIN-SERO & 88.73±1.36 & 88.69±1.41 & 89.99±1.71 & 86.29±1.74 & 86.80±0.90\\
STAGIN-GARO & 88.34±0.94 & 88.33±0.91 & 89.50±2.90 & 86.76±4.73 & 97.18±0.13\\
FBNETGNN & 88.05±0.15 & 87.93±0.97 & 89.60±1.21 & 86.11±1.32 & 86.51±0.89\\
\hline 
\textit{\textbf {Machine Learning Model}}\\
SVM & 87.55±1.79 & 87.53±1.91 & 90.16±2.31 & 84.31±2.05 &  85.76±2.07\\
Logistic Regression & 92.16±0.77 & 92.10±0.71 & 93.18±1.17 & 90.91±2.04 & 91.16±0.84 \\
Forest & 77.16±2.63 & 77.42±2.64 & 85.23±2.05 & 67.09±4.21 & 72.29±3.43 \\
\hline
\textit{\textbf {Ours}}\\
\rowcolor{Gray}
\bf SG-BrainMAE & \bf 97.49±0.15 & \bf 97.46±0.18 & \bf 97.66±0.91 & \bf 97.28±1.19 &  \bf 97.18±0.13 \\
\rowcolor{Gray}
\bf AG-BrainMAE & \bf 97.13±0.56 & \bf 97.17±0.61 & \bf 97.62±0.95 & \bf 96.14±0.76 &  \bf 96.76±0.61\\
\bottomrule
\end{tabular}}
}
\end{table*}
\begin{table*}[h]
\setlength{\tabcolsep}{4 pt}
\renewcommand{\arraystretch}{1.2}
\centering
\caption{HCP-3T behavior prediction}
\label{appdx-tbl-all-hcp3t-beh}
\resizebox{1\linewidth}{!}{
{\begin{tabular}{llllllllllll}
\toprule
\multirow{2.5}{*}{Model}&\multicolumn{8}{c}{Behaviors (measured in MAE)}\\
\cmidrule(lr){2-9} 
&{PicSeq}&{PMT$\_$CR}&{PMT$\_$SI}&{PicVocab}&{IWRD}&{ListSort}&{LifeSatisf}&{PSQI} \
\\ \hline 
\textit{\textbf {FIX-FC}}\\
BrainNetTF & 7.11±0.22 & 2.28±0.11 & 1.72±0.12 & 4.70±0.10 & 1.56±0.03 & 5.96±0.07& 4.96±0.22& 1.49±0.05\\
VanillaTF & 8.19±0.46 & 2.73±0.07 & 2.13±0.07 & 5.93±0.14 & 1.81±0.08 & 6.91±0.21 & 5.71±0.08 & 1.68±0.05 \\
BrainNetCNN & 10.21±0.22 & 3.25±0.13 & 2.64±0.14 & 6.65±0.27 & 2.26±0.05 & 8.51±0.20 & 7.12±0.24 & 1.68±0.05 \\
 \hline 
\textit{\textbf {Dynamic-FC}}\\
STAGIN-SERO & 10.22±0.15 & 3.49±0.05 & 2.70±0.06& 6.78±0.20 & 2.26±0.05 & 8.51±0.20 & 7.12±0.24 & 2.12±0.04\\
STAGIN-GARO & 10.26±0.18 & 3.44±0.10 & 2.69±0.09 & 6.92±0.30 & 2.25±0.04 & 8.52±0.26 & 7.09±0.35 & 2.08±0.04 \\
FBNETGNN & 8.62±0.21 & 2.93±0.11 & 2.34±0.11 & 5.83±0.15 & 1.96±0.04 & 7.31±0.10 & 6.09±0.10 & 1.81±0.03 \\
\hline 
\textit{\textbf {ML Model}}\\
Oridinary Regression & 11.23±0.25 & 4.05±0.07 & 3.34±0.07 & 7.50±0.13 & 2.58±0.08 & 9.90±0.26 & 8.05±0.18 & 2.50±0.07\\
Ridge & 8.91±0.23 & 3.18±0.10 & 2.63±0.10 & 6.05±0.10 & 2.05±0.06 & 7.70±0.23 & 6.42±0.02 & 2.01±0.09 \\
ElasticNet & 10.83±0.14 & 4.01±0.03 & 3.29±0.04 & 7.44±0.22 & 2.35±0.05 & 9.21±0.19 & 7.29±0.28 & 2.15±0.07 \\
\hline
\textit{\textbf {Ours}}\\
\rowcolor{Gray}
\bf SG-BrainMAE & \bf 5.06±0.21 & \bf 1.63±0.08 & \bf 1.24±0.04 & \bf 3.40±0.14 & \bf 1.11±0.04 & \bf 4.35±0.12 & \bf 3.64±0.27 & \bf 1.05±0.06\\
\rowcolor{Gray}
\bf AG-BrainMAE & \bf 5.09±0.05 & \bf 1.67±0.10 & \bf 1.28±0.06 & \bf 3.34±0.11 & \bf 1.13±0.03 & \bf 4.37±0.06 & \bf 3.58±0.17 & \bf 1.07±0.05 \\
\bottomrule
\end{tabular}}
}
\end{table*}
\begin{table*}[t]
\centering
\setlength{\tabcolsep}{4 pt}
\renewcommand{\arraystretch}{1.2}
\caption{HCP-Aging age prediction}
\label{appdx-tbl-all-hcpaging}
{
\resizebox{0.9\linewidth}{!}{
\begin{tabular}{llllllll}
\toprule
\multirow{2.5}{*}{ Model}&\multicolumn{5}{c}{ Gender}&{}&\multirow{2.5}{*}{ Aging(MAE)}\\
\cmidrule(lr){2-6} 
&{Accuracy(\%)}&{AUROC(\%)}&{specificity(\%)}&{sensitivity(\%)}&{F1 Score(\%)}
\\ \hline
\textit{\textbf {FIX-FC}}\\
BrainNetTF & 90.21±3.81 & 90.73±2.85 & 87.97±9.26 & 91.84±6.56 & 89.35±3.49 &{}& 6.15±0.71 \\
VanillaTF & 88.96±2.16 & 88.76±2.21 & 89.22±2.30 & 88.56±3.01 & 87.54±2.53 &{}& 6.78±0.56  \\
BrainNetCNN & 88.83±1.52 & 88.74±1.58 & 89.87±2.85 & 87.92±3.18 & 87.48±1.20 &{}& 8.71±0.62  \\
 \hline 
\textit{\textbf {Dynamic-FC}}\\
STAGIN-SERO & 82.37±1.66 & 82.57±1.36 & 85.23±4.68 & 76.00±6.51 & 77.91±3.14 &{}& 8.96±0.47 \\
STAGIN-GARO & 80.67±0.81 & 80.58±1.03 & 84.63±4.05 & 77.95±4.43 & 78.81±3.27 &{}& 8.65±0.28  \\
FBNETGNN & 89.50±3.58 & 89.34±3.49 & 90.04±1.94 & 89.05±5.18 & 88.35±3.45 &{}& 6.68±1.00 \\
\hline 
\textit{\textbf {ML Model}}\\
SVM & 86.04±0.97 & 86.23±0.98 & 90.87±0.66 & 79.84±1.76 & 83.36±1.48 &{}&- \\
Logistic Regression & 89.49±1.27 & 89.41±1.23 & 90.92±0.81 & 87.66±2.17 & 88.20±1.50&{}&- \\
Forest & 73.75±1.44 & 74.42±0.94 & 85.82±1.46 & 58.35±4.11 & 66.05±2.47 &{}&- \\
OrdinaryRegression &-&-&-&-&-&{}&7.63±0.21\\
Ridge &-&-&-&-&-&{}&  7.08±0.20 \\
ElasticNet &-&-&-&-&-&{}& 9.43±0.61\\
\hline 
\textit{\textbf {Ours}}\\
\rowcolor{Gray}
\bf SG-BrainMAE & \bf 92.67±1.07 & \bf 92.51±1.07 & \bf 97.66±0.91 & \bf 97.28±1.19 &  \bf 97.18±0.13 &{}& \bf 5.78±0.44 \\
\rowcolor{Gray}
\bf AG-BrainMAE & \bf 91.12±1.99 & \bf 91.15±2.03 & \bf 97.62±0.95 & \bf 96.14±0.76 &  \bf 96.76±0.61 &{}& \bf 6.49±1.00 \\
\bottomrule
\end{tabular}}
}
\end{table*}
\begin{table*}[t]
\setlength{\tabcolsep}{4 pt}
\renewcommand{\arraystretch}{1.2}
\centering
\caption{NSD Task performance prediction}
\label{appdx-tbl-all-nsd}
\resizebox{0.4\linewidth}{!}
{\begin{tabular}{lll}
\toprule
\multirow{2.5}{*}{Model}&\multirow{2.5}{*}{ Task Accuracy}&\multirow{2.5}{*}{ RT(ms)}\\
\addlinespace
\\ \hline 
\textit{\textbf {FIX-FC}}\\
BrainNetTF & 0.070±0.003 & 92.344±2.343  \\
VanillaTF & 0.075±0.004 & 96.252±2.133   \\
BrainNetCNN & 0.078±0.004 & 102.911±2.225 \\
 \hline 
\textit{\textbf {Dynamic-FC}}\\
STAGIN-SERO & 0.089±0.003 & 116.635±2.197 \\
STAGIN-GARO & 0.091±0.002 & 116.130±2.099   \\
FBNETGNN & 0.074±0.005 & 95.349±2.320  \\
\hline 
\textit{\textbf {ML Model}}\\
Ordinary Regression & 0.117±0.003 & 157.460±3.788  \\ 
Ridge & 0.093±0.003 & 129.609±2.518 \\
ElasticNet & 0.110±0.004 & 174.946±2.585 \\
\hline 
\textit{\textbf {Ours}}\\
\rowcolor{Gray}
\bf SG-BrainMAE & \bf 0.069±0.004 & \bf 90.678±1.767  \\
\rowcolor{Gray}
\bf AG-BrainMAE & \bf 0.070±0.004 & \bf 92.154±2.265 \\
\bottomrule
\end{tabular}}
\end{table*}

\begin{table*}
\centering
\caption{Behaviors description}
\label{appdx-tbl-beh}
\resizebox{1\textwidth}{!}{
\begin{tabularx}{\linewidth}{p{18mm}p{40mm}p{70mm}}
\toprule
\addlinespace
Behavior & Display Name & Description\\
\addlinespace
\hline 
\addlinespace
PicSeq & NIH Toolbox Picture Sequence Memory Test: Unadjusted Scale Score & The Picture Sequence Memory Test is a measure developed for the assessment of episodic memory for ages 3-85 years. Participants are given credit for each adjacent pair of pictures,the maximum score is 17(because that is the number of adjacent pairs of pictures that exist).  \\
\addlinespace
\hline
\addlinespace
\text{PMT\_CR} & Penn Progressive Matrices: Number of Correct Responses \text{(PMAT24\_A\_CR)} & Penn Matrix Test: Number of Correct Responses. A measure of abstraction and mental flexibility. It is a multiple choice task in which the participant must conceptualize spatial, design and numerical relations that range in difficulty from very easy to increasingly complex. \\
\addlinespace
\hline
\addlinespace
\text{PMT\_SI} & Penn Progressive Matrices: Total Skipped Items \text{(PMAT24\_A\_SI)} & Penn Matrix Test: Total Skipped Items (items not presented because maximum errors allowed reached).\\
\addlinespace
\hline
\addlinespace
PicVocab & NIH Toolbox Picture Vocabulary Test: Unadjusted Scale Score & This measure of receptive vocabulary is administered in a computerized adaptive format. The respondent is presented with an audio recording of a word and four photographic images on the computer screen and is asked to select the picture that most closely matches the meaning of the word. \\
\addlinespace
\hline
\addlinespace
IWRD & Penn Word Memory Test:  Total Number of Correct Responses \text{(IWRD\_TOT)} &  Participants are shown 20 words and asked to remember them for a subsequent memory test. They are then shown 40 words (the 20 previously presented words and 20 new words matched on memory related characteristics).\\
\addlinespace
\hline
\addlinespace
ListSort & NIH Toolbox List Sorting Working Memory Test: Unadjusted Scale Score  & This task assesses working memory and requires the participant to sequence different visually- and orally-presented stimuli. Pictures of different foods and animals are displayed with both a sound clip and written text that name the item.  Participants are required to order a series of objects (either food or animals) in size order from smallest to largest.   \\
\addlinespace
\hline
\addlinespace
LifeSatisf & NIH Toolbox General Life Satisfaction Survey: Unadjusted Scale Score  & Life Satisfaction is a concept within the Psychological Well-Being subdomain of Emotion. Life Satisfaction is one's cognitive evaluation of life experiences and is concerned with whether people like their lives or not. This self-report measure is a 10-item calibrated scale comprised of items from the Satisfaction with Life Scale.  \\
\addlinespace
\hline
\addlinespace
PSQI & Sleep (Pittsburgh Sleep Questionnaire) Total Score & The Pittsburgh Sleep Quality Index (PSQI) is a self-rated questionnaire which assesses sleep quality and disturbances over a 1-month time interval.Scores for each question range from 0 to
3, with higher scores indicating more acute sleep
disturbances.\\
\addlinespace
\bottomrule
\end{tabularx}
}
\end{table*}

\clearpage
\section{Additional fMRI Signal Reconstruction Results By BrainMAE}
\label{appdx-more_recon}
\begin{figure*}[h]
\begin{center}
\includegraphics[width=1\textwidth]{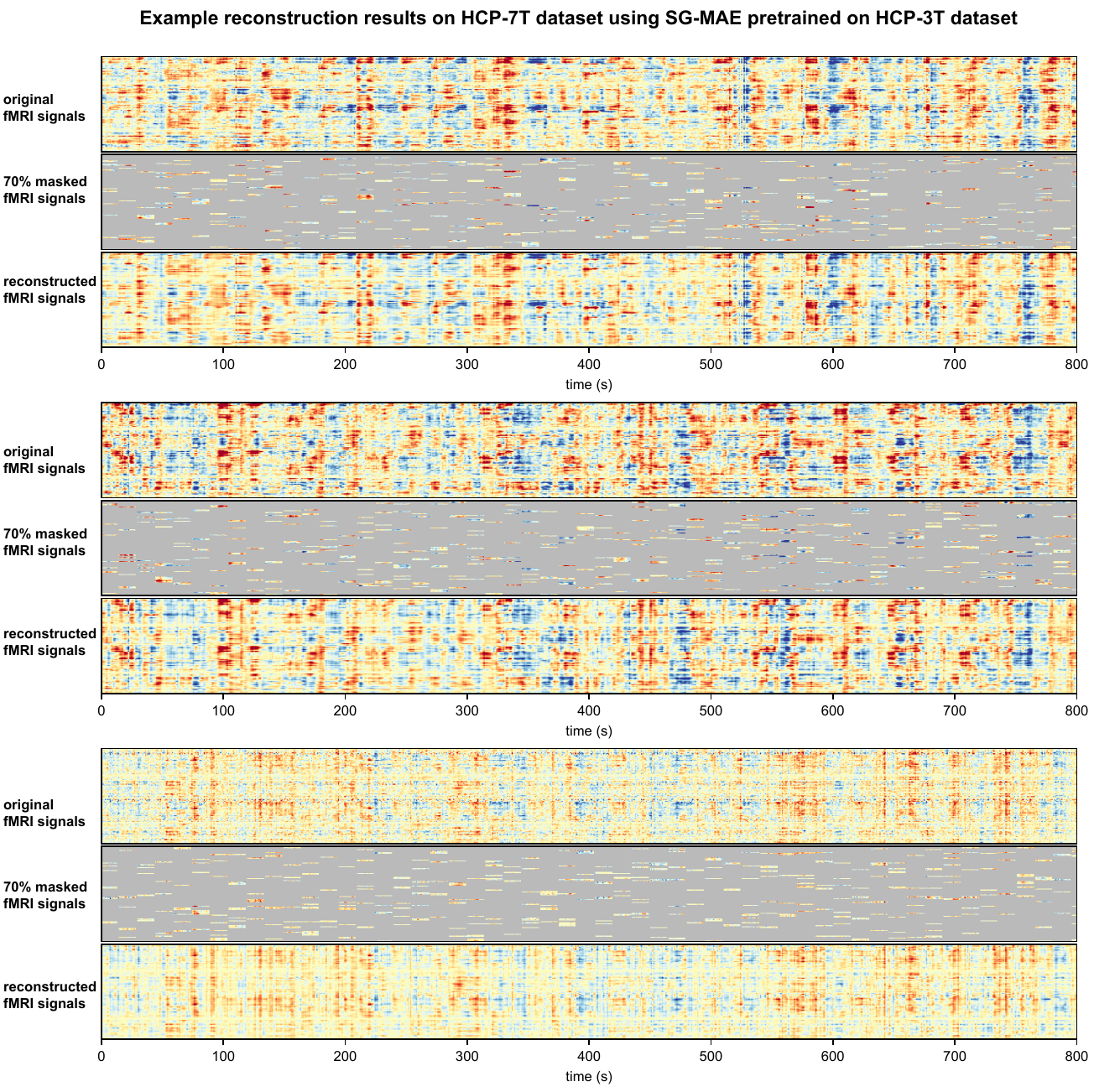}
\end{center}
\caption{Example reconstruction results on HCP-7T using HCP-3T pretrained model.}
\label{fig-S2}
\end{figure*}

\begin{figure*}[h]
\begin{center}
\includegraphics[width=1\textwidth]{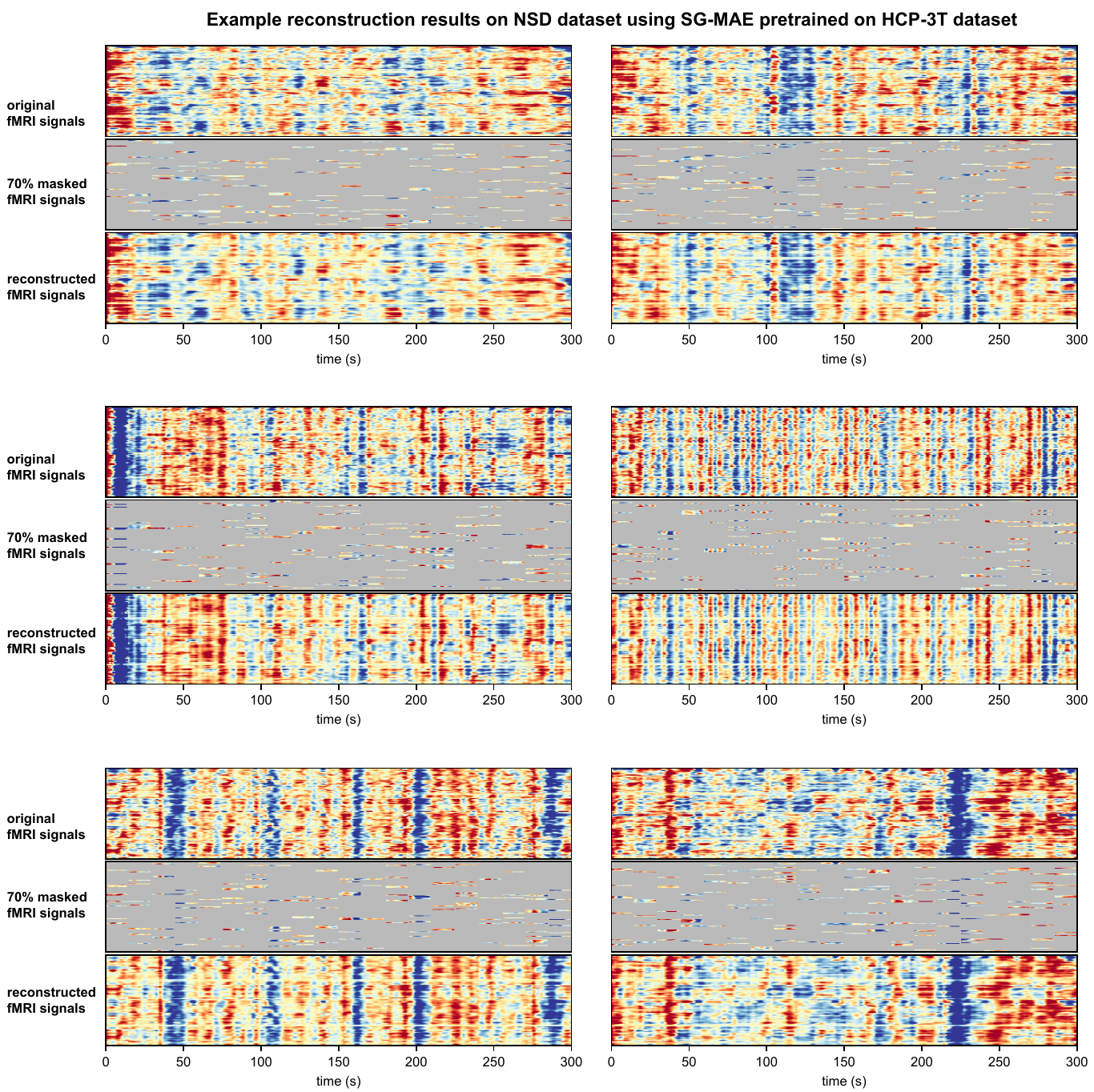}
\end{center}
\caption{Example reconstruction results on NSD using HCP-3T pretrained model.}
\label{fig-S3}
\end{figure*}

\begin{figure*}[h]
\begin{center}
\includegraphics[width=1\textwidth]{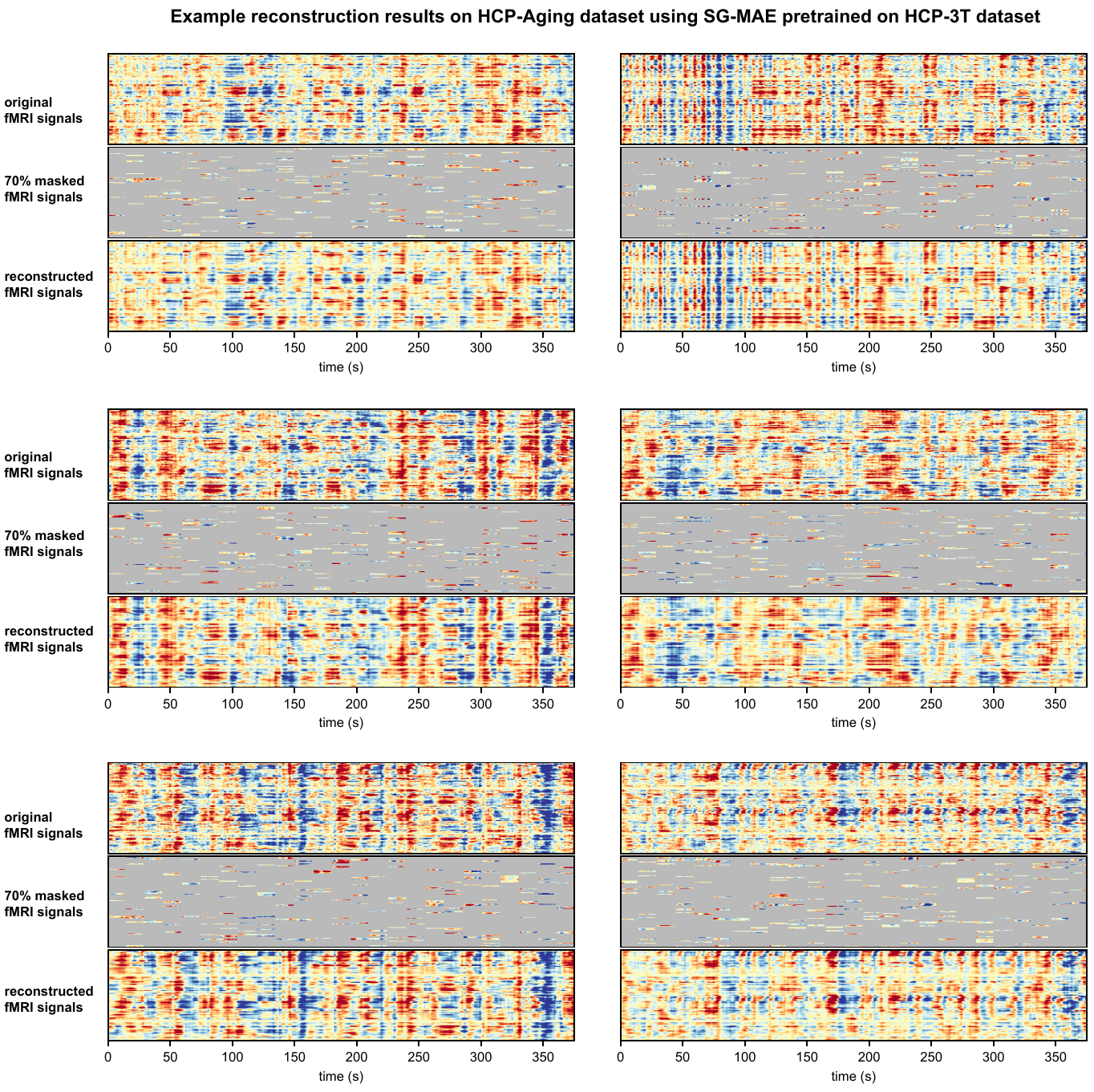}
\end{center}
\caption{Example reconstruction results on HCP-Aging dataset using HCP-3T pretrained model.}
\label{fig-S4}
\end{figure*}

Additional reconstruction examples of applying the HCP-3T pre-trained SG-BrainMAE to various unseen fMRI datasets are shown in Figure~\ref{fig-S2} for the HCP-7T dataset, Figure~\ref{fig-S3} for the NSD Task dataset, and Figure~\ref{fig-S4} for the HCP-Aging dataset. These examples demonstrate the generalizable representation learned by BrainMAE.

\end{document}